\documentstyle[prl,aps,preprint,tighten]{revtex}
\begin{document}
\pagenumbering{arabic}
\title{ Josephson and Quasiparticle Tunneling between Anisotropically Paired
Superconductors}
\author{Yury S. Barash, and Anatoly A. Svidzinsky \\  }
\address {I.E. Tamm Department of Theoretical Physics, P.N. Lebedev Physics
Institute \\
Leninsky Prospect 53, Moscow 117924, Russia \\ }
\maketitle

\begin{abstract}
Charge transport in tunnel junctions between singlet anisotropically paired
superconductors is investigated theoretically making use of the Eilenberger
equations for the quasiclassical Green functions. For specularly
reflecting tunnel barrier plane we have found and described
analytically characteristic singular points of the I-V curves which are
specific for the case of anisotropic pairing. All four terms for the electric
current are examined. Two of them describe ac Josephson effect and two
correspond to the quasiparticle current (the last term occurs only for a
time-dependent voltage). Various momentum dependences of the order parameter in
the bulk of superconductors and different crystal orientations are considered.
The results of numerical calculations for the total I-V curves are presented
for some particular cases.
\end{abstract}
\section{Introduction}

Theoretical description of the Josephson and quasiparticle currents through
a tunnel junction was developed many years ago for the case of $s$-wave
isotropic superconductors. In particular, the total tunnel
current under the externally applied time-dependent voltage was investigated
in detail microscopically making use of tunneling Hamiltonian method
\cite{lar,wer} (see also \cite{bar} and references therein).
Now it becomes clear that in the case of anisotropically paired superconductors
the measurements of the Josephson effect and the quasiparticle current provide
an important information about the structure of the superconducting
order parameters from both sides of the junction and about the proximity and
some other specific surface superconducting effects at the tunnel barrier plane
\cite{gesh,sig1,mil,b2,buch2,sig2}. The investigations in this field are
of great interest and Josephson and quasiparticle tunneling experiments
studying the anisotropic structure of superconducting order parameter for
various high-temperature superconductors have attracted now much attention
(see, for example,\cite{har,renfisch}).

One of the characteristic peculiarities of microscopic description
of charge transport through tunnel junctions in anisotropically paired
superconductors is, in fact, the failure of the tunneling Hamiltonian
method applied to this case \cite{b1}. This approach,
being usually quite suitable for tunnel junctions between $s$-wave isotropic
superconductors, contains ambiguities in the case of anisotropic pairing due to
substantial momentum dependence of the matrix elements describing tunneling
between superconductors. One can show that the choice of the tunneling matrix
elements as being independent on the momentum direction (standard for the
s-wave case) leads to confusing results for anisotropically paired
superconductors. Furthermore, an unambiguous choice of this dependence cannot
be done within this method. This leads to the necessity of making use of a
microscopic description of tunneling between such superconductors based on
matching of the electron propagators at the tunnel barrier, since this approach
leaves no space for ambiguity.

Below we develop a microscopic study of the singular points  on
current-voltage characteristics for ac Josephson effect and for
quasiparticle
current of tunnel junctions between anisotropically paired superconductors
under the externally applied ( and, generally speaking, time-dependent) voltage.
In our microscopic approach we are based on the Eilenberger equations for
quasiclassical electron propagators combined with corresponding boundary
conditions and the microscopic expression for the tunnel current.
In the particular case of $s$-wave isotropic superconductors our results
coincide with those found by Larkin and Ovchinnikov \cite{lar}. They obtained
general expressions and described the singular points for functions $I_m(V)$
($m=1,2,3,4$), which enter the relation for the total tunnel current (see
below, for example, Eqs.(\ref{cur5})-(\ref{jm})). Here $I_{1,2}(V)$ describe
the amplitudes of two terms in the expression for the Josephson current, which
are reduced for time-independent voltage to $I_{1}(V)\sin \left( \chi _1-\chi
_2+2eVt/\hbar\right)+ I_{2}(V)\cos \left( \chi _1-\chi _2+2eVt/\hbar\right)$.
Further, $I_{3,4}(V)$ describe the quasiparticle current. The function $I_4(V)$
occurs only in the case of time-dependent voltage while for permanent voltage
the quasiparticle current is entirely reduced to the only term $I_3(V)$.  It
was shown, in particular, that at $|eV|=\Delta_1+\Delta_2$ the singular parts
of functions $I_{1,4}(V)$ diverge logarithmically, while the functions
$I_{2,3}(V)$ undergo a jump. Analogous singularities were found for
time-dependent voltage oscillating with the frequency $\hbar\omega=\Delta_1+
\Delta_2$.
The singularity of $I_{1}$ is known as Riedel singularity \cite{ried}.
It is closely associated with the corresponding singularity in the
density of states for superconductors at $\hbar\omega=\Delta$.

It is clear that Riedel's anomalies themselves must be
washed out in the case of strongly anisotropic pairing when
the density of states doesn't have the divergence.
Nevertheless, as we show below, some new characteristic singularities
appear even in the presence of nodes in the order
parameter on the Fermi surface.
The characteristic behaviors of the I-V curves turn out
to be strongly dependent upon the crystal orientations of the superconductors,
which, in particular, govern the spatial dependence of the order parameters
near the barrier plane.
But even in the simplest case of uniform distribution of the superconducting
order parameter on both sides of a tunnel barrier the anomalies
become strongly modified as compared to the isotropic case.
The point is that for anisotropically paired
singlet superconductors the functions $I_m(V)$ become also depending upon
the momentum direction on the Fermi surface and enter the expression for the
tunnel current as corresponding integrands.
For obtaining the current-voltage
relation in the vicinities of singular points one should carry out the
integration over the momentum directions. We show that the nonanalytical
behavior of the I-V curves takes place only for the voltages
$|eV|$ which are equal to the  values of the expressions
$\left| |\Delta _2(\hat{{\bf p}}_2)| \pm |\Delta _1(\hat{{\bf p}}_1)|\right|$
at their extremal points. Here $\hat{{\bf p}}_1$ is the direction of the
incident momentum and $\hat{{\bf p}}_2$ -- of the transmitted momentum,
which is directly connected with $\hat{{\bf p}}_1$ and the shapes of the Fermi
surfaces. All the expressions are considered, for example, as functions of
$\hat{{\bf p}}_1$. Further, the singular behaviors of the I-V curves
at these extrema turn out to be strongly dependent upon the type of the
extremal points. This was pointed out earlier in \cite{b2}, where
the characteristic behaviors of the I-V curves have been determined for the
important particular case of the quasiparticle current at low temperatures,
for the crystalline orientations when there is no surface pair breaking.

The situation becomes more complicated for the crystal orientations, for
which the spatial dependence of the order parameter near the barrier plane
plays an important role. The possibility for existence of quasiparticle
bound states located in the vicinity of barrier plane is of importance in
this case. At least two factors may be associated with such kind of effects.
The former is the zero energy quasiparticle state which occurs at the barrier
plane under certain general conditions \cite{buch2,h,yh,mat2,nag}. The latter
is the appearance of additional quasiparticle bound surface states due to the
spatial dependence of the order parameter, which is suppressed by the surface
\cite{buch2}. We discuss below these possibilities and respective
consequences for the current-voltage characteristics.

\section{Microscopic Expression for the Tunnel Electric Current}

We consider a tunnel junction with transparency coefficient $D\ll 1 $ and
specularly reflecting barrier plane between two clean singlet
superconductors. The external voltage $V(t)=\Phi_2(t)-\Phi_1(t)$ is assumed
to be applied to the junction. Let the normal to the junction barrier plane
be directed along the $x$-axis: ${\bf n}\parallel Ox$. Then the microscopic
expression for a tunnel current density in the first order of the barrier
transparency $D$ may be represented in the form \cite{zai}:
$$
j_x=-\frac {\displaystyle 1}{\displaystyle 8\pi^3 }\int\limits_{v_{x1}>0}
\frac{\displaystyle d^2S_1\cdot v_{x1}}{\displaystyle 4\pi \cdot v_{f1}}Sp \
\hat \tau_3 D\left(\hat{{\bf p}}_1\right) \left( \hat g_1^R\cdot \hat
g_2^K+\hat g_1^K\cdot \hat g_2^A - \qquad \qquad \qquad \right.
$$
\begin{equation}
\label{cur1}\left. \qquad \qquad \qquad \qquad \qquad \qquad \qquad \qquad
-\hat g_2^R\cdot \hat g_1^K-\hat g_2^K \cdot \hat g_1^A\right) _{\circ
}\left( t,t\right) \enspace.
\end{equation}
Here and below we put $e=1$ as well as $\hbar,c=1$.

The retarded, advanced and Keldysh quasiclassical matrix propagators are
taken in Eq.(\ref{cur1}) at the junction barrier plane and must be
calculated in zeroth order in junction transparency (i.e. for impenetrable
barrier). They depend upon the respective momentum directions $\hat{{\bf p}}
_1$ and $\hat{{\bf p}}_2$. Index $1\,(2)$ labels the left (right) half space
with respect to the boundary plane, and $v_x$ is the Fermi velocity
component along the normal to the plane interface ${\bf n}$. The integration
in Eq.(\ref{cur1}) is carried out over the part of the Fermi surface with $
v_x>0$. The relation between the incident and transmitted Fermi momenta (
that is between ${\bf p}_1$ and ${\bf p}_2$ ) is as follows. The components
parallel to the specular plane interface are equal to each other, while the
values of normal components are determined by the condition, that ${\bf p}_1$
and ${\bf p}_2$ lie on respective Fermi surfaces. Naturally, in the
particular case of identical superconductors with the spherical or cylindrical
Fermi surfaces (provided the cylindrical axes are parallel to the barrier plane)
the total incident and transmitted momenta are equal to each other
${\bf p}_1={\bf p}_2$.

Notations used in Eq.(\ref{cur1}) are in correspondence with the following
example:
\begin{equation}
\label{ex1}(\hat g_1^R\cdot \hat g_2^K)_{\circ }\left( t,t\right)
=\int\limits_{-\infty }^tdt_1\hat g_1^R\left( t,t_1\right) \cdot \hat
g_2^K\left( t_1, t\right) \enspace .
\end{equation}

Here $\hat g_l^{R,A,K}\left( t,t_1\right) =\hat S_l\left( t\right) \hat
g_l^{R,A,K}\left( t-t_1 \right)\hat S_l^{+}\left( t_1\right) $\, ($l=1,2$),
where

\begin{equation}
\label{S}\hat S_l\left( t\right) =\left(
\begin{array}{cc}
e^{i\chi_l \left( t\right)/2} & 0 \\
0 & e^{-i\chi_l \left( t\right)/2}
\end{array}
\right) \enspace , \enspace
\chi_l \left( t\right) =\chi_l -2\int\limits^t\Phi_l \left( t^{\prime
}\right) dt^{\prime }\enspace
\end{equation}
and $\chi_l$ is the phase of the order parameter of $l$-th superconductor at
the boundary plane for the case of zero electric potential $\Phi_l$.

Below we consider only singlet types of anisotropic pairing, for which
matrix propagators may be represented in the form
\begin{equation}
\label{hatg}\hat g = \left(
\begin{array}{cc}
g & f \\
f^+ & -g
\end{array}
\right) \enspace .
\end{equation}

For further calculations it is also important that nonequilibrium effects,
as a rule, are not essential in tunnel junctions. The voltage $V$ simply
shifts the Fermi levels in the electrodes relative each other by the value $
V $. Besides, for finding the current in the first approximation with regard
to transparency, the calculation of the Green functions may be performed for
impenetrable half spaces (disregarding the transmissions of electrons
through the junction). Under the conditions we consider the distribution
functions of electrons remain the equilibrium ones and the effect of
electric potential results only in the appearance of corresponding spatial
independent terms in phases of superconducting order parameters and Green
functions on both banks of the junction (see, for instance, Eq.(\ref{S})).
Then the following relationship is valid

\begin{equation}
\label{rel}\hat g^K(\omega)=\left(\hat g^R(\omega)-\hat g^A(\omega)\right)
\tanh\left(\frac{\omega}{2T}\right),
\end{equation}

According to general symmetries of the propagators one has
$\hat g^A= \hat \tau_3\left( \hat g^R\right)^\dagger\hat\tau_3$,
where $\hat \tau_3$ is the third Pauli matrix.
We suppose that the complex phase of the order parameter within the
superconducting half space with impenetrable boundary is constant in the
coordinate and momentum spaces. Then this phase is just the quantity $\chi_l$
in the Eq.(\ref{S}) and the Green functions further must be calculated for
the real values of the order parameter, though its sign may depend on the
momentum direction. Under these conditions one has
$f^{R,A}(-\hat{{\bf p}},x,\omega)=-f^{+R,A}(\hat{{\bf p}},x,\omega)$,
$g^{R,A}(-\hat{{\bf p}},x,\omega)=g^{R,A}(\hat{{\bf p}},x,\omega)$,
$f^{R,A}(\hat{{\bf p}},x,-\omega)=f^{R,A*}(\hat{{\bf p}},x,\omega)$,
$g^{R,A}(\hat{{\bf p}},x,-\omega)=-g^{R,A*}(\hat{{\bf p}},x,\omega)$.
 We take into account as well that a propagator taken at
the impenetrable boundary for the incoming momentum $\hat{{\bf p}}$ is equal
to the one taken for the outgoing momentum $\check{{\bf p}}$, and $v_x(\hat{
{\bf p}})=-v_x(\check{{\bf p}})$. Then one can get that for the
time-dependent voltage $V(t)=V_0+a\cos (\omega_0 t)$ the expression (\ref
{cur1}) for the tunnel current acquires the form

$$
j_x=\sum_{n=-\infty }^\infty J_n\left( \frac a{\omega _0}\right)
\left\{j_1(V_0+\omega _n)\sin \left( \chi _1-\chi _2+2V_0t+\frac a{\omega
_0}\sin \left( \omega _0t\right) +\omega _nt\right)+ \right.
$$
$$
\qquad\qquad\qquad\qquad\quad +j_2(V_0+\omega _n)\cos \left( \chi _1-\chi
_2+2V_0t+ \frac a{\omega _0}\sin \left( \omega _0t\right) +\omega
_nt\right)+
$$

\begin{equation}
\label{cur5}\left. \!+j_3(V_0+\omega _n)\cos \left(\!\omega _nt- \frac
a{\omega _0}\sin\left(\omega _0t\right) \!\right)+j_4(V_0+\omega
_n)\sin\left(\!\omega _nt- \frac a{\omega _0}\sin\left(\omega
_0t\right)\!\right)\!\right\}.
\end{equation}

The following notations are introduced here

\begin{equation}
\label{jm}j_m(V)=\int\limits_{v_{x1}>0}\frac{d^2S_1}{(2\pi )^3}\frac{v_{x1}}{
v_{f1}}D\left(\hat{{\bf p}}_1\right) I_m(V,\hat{{\bf p}}_1) \enspace,
m=1,2,3,4 \enspace,
\end{equation}
\begin{equation}
\label{i1}I_1(V,\hat {{\bf p}}_1)=\!\!\int\limits_{-\infty }^\infty \!\!\frac{
d\omega }{2\pi ^2}\tanh \left( \frac \omega {2T}\right) Im\left( f_1^R\left(
\omega -V\right) \left( f_2^{+R}\left( \omega \right) +f_2^{R*}\left( \omega
\right) \right) +\left( 1\leftrightarrow 2\right) \right)\!\! \enspace ,
\end{equation}
\begin{equation}
\label{i2}I_2(V,\hat {{\bf p}}_1)=\!-\!\!\!\int\limits_{-\infty }^\infty
\!\!\!\frac{ d\omega }{2\pi ^2}\tanh \left(\! \frac \omega {2T}\!\right)
Re\left( f_1^R\left( \omega -V\right) \left( f_2^{+R}\left( \omega \right)
+f_2^{R*}\left( \omega \right) \right) +\left( 1\leftrightarrow 2\right)
\right)\!\! \enspace ,
\end{equation}
\begin{equation}
\label{i3}I_3(V,\hat{{\bf p}}_1)=\int\limits_{-\infty }^\infty \frac{d\omega
}{\pi^2} \left( \tanh \left( \frac{\omega -V}{2T}\right) -\tanh \left( \frac
\omega {2T}\right) \right) Img_1^R\left( \omega \right) \cdot Img_2^R\left(
\omega -V\right) \enspace ,
\end{equation}
\begin{equation}
\label{i4}I_4(V,\hat {{\bf p}}_1)\!=\!-\!\!\!\!\int
\limits_{-\infty }^\infty \!\!\!
\frac{d\omega }{\pi ^2}\tanh \left( \frac \omega {2T}\right)\!\! \left(
Reg_1^R\left( \omega \!-\!V\right)  Img_2^R\left( \omega \right)
\!+\!Reg_2^R\left( \omega\! -\!V\right) Img_1^R\left( \omega \right)
\right)\!.
\end{equation}

The quantities $I_{1,4}(V)$ are even functions on $V$ and $I_{2,3}$ - odd
functions. Note, that the interchanges $f\leftrightarrow f^+ $ in the
expressions (\ref{i1}), (\ref{i2}) lead to the same result for the Josephson
current after the integration over the momentum directions in (\ref{jm}).
Eqs.(\ref{cur5})-(\ref{i4}) are in correspondence with those obtained by
Larkin and Ovchinnikov
\cite{lar} in consideration of s-wave isotropic superconductors.

In the case of permanent voltage through the junction one should let $a=0$
in Eq.(\ref{cur5}). Then one gets
\begin{equation}
\label{curp}j_x(V)=j_1(V)\sin \left( \chi _1-\chi _2+2Vt\right) +j_2(V)\cos
\left( \chi _1-\chi _2+2Vt\right) +j_3(V) \enspace .
\end{equation}

It follows from Eqs.(\ref{curp}) and (\ref{cur5}), that the specific
singular points on the voltage-current characteristic for the tunnel current
for both time-dependent and permanent voltages through the junction are
defined by the singular points of functions $j_{m}(V)$. If corresponding
singular points are $V=V_{m}$, then in the case of permanent voltage the
specific points on the current-voltage characteristics are $V_{1,2,3}$ \, ,
while for the current (\ref{cur5}) peculiarities appear for the values $
V_{m,n}=V_m-n\omega_0$\ \ ($m=1,2,3,4;\ n=0, \pm 1, \pm 2, \ldots$). For
large enough values $n$ ($ (a/\omega_0)\lesssim n$) the amplitude of the
current becomes quite small (according to respective behavior of the Bessel
functions).

\section{Quasiclassical Green Functions at the Impenetrable Boundary Plane}

As it follows from previous section, for the calculation of the tunnel
electric current one should consider the retarded electron propagators for
the impenetrable superconducting half space and then find the values of the
propagators at the boundary plane. The total analytical solution of this
problem is quite complicated and hasn't yet been obtained for any particular
pairing potential leading to an anisotropically paired superconductivity (
excluding the particular orientation for which there is no surface pair
breaking and the order parameter doesn't manifest a spatial dependence). The
problem may be essentially simplificated if one is interested only in the
singular points on the I-V curves and consequently the singular points of
the propagators taken at the barrier plane. For the consideration of this
problem we make use of the Eilenberger equations for the retarded
quasiclassical propagators, which may be written for the case of
superconductors with singlet pairing as follows (further we omit the
superscript for the retarded propagators)

\begin{equation}
\label{ee}\left\{
\begin{array}{l}
(2\omega +iv_{x} \partial _x)f(
\hat{{\bf p}},x,\omega )+2\Delta (\hat{{\bf p}},x) g(\hat{{\bf p}},x,\omega
)=0 \\ (2\omega -iv_{x} \partial _x)f^{+}(
\hat{{\bf p}},x,\omega )- 2\Delta ^{*}(\hat{{\bf p}},x)g(\hat{{\bf p}}
,x,\omega )=0 \\ iv_{x} \partial _x g(\hat{{\bf p}},x,\omega )-\Delta (\hat{
{\bf p}},x) f^{+}(\hat{{\bf p}},x,\omega)-\Delta ^{*}(\hat{{\bf p}},x) f(
\hat{{\bf p}},x,\omega )=0 \enspace .
\end{array}
\right.
\end{equation}

For the sake of definiteness let us assume that the superconductor occupies
a half space $x>0$.

Apart from a self-consistency equation for $\Delta (\hat{{\bf p}},x)$,
Eqs.(\ref{ee}) have to be supplemented by a normalization condition

\begin{equation}
\label{n}g^2+f\cdot f^{+}=-\pi^{2} \enspace ,
\end{equation}
and boundary conditions for quasiclassical propagators. For the specularly
reflecting impenetrable surface one gets

\begin{equation}
\label{bc0}g(\hat {{\bf p}},\omega )=g(\check {{\bf p}},\omega )\mid _{x=0}
\enspace ,\enspace f(\hat {{\bf p}},\omega )=f(\check {{\bf p}},\omega )\mid
_{x=0}\enspace ,\enspace f^{+}(\hat {{\bf p}},\omega )=f^{+}(\check {{\bf p}
},\omega )\mid _{x=0}\enspace \enspace .
\end{equation}
Here $\hat {{\bf p}}$ is the direction of incident momentum and $\check {
{\bf p}}$ - the direction of reflected momentum.

The behaviors of the propagators in the depth of the superconductor

\begin{equation}
\label{bcd}g(\hat {{\bf p}},\omega )\!\mid _{x=\infty }\!=\!\frac{-\pi\omega
}{\sqrt{ |\Delta _\infty (\hat {{\bf p}})|^2-\omega ^2}}\! \enspace\! ,\!
\enspace \! f\mid _{x=\infty }=-f^{+}\left( \hat {{\bf p}},\omega \right)\!
\mid _{x=\infty }\!=\!\frac{\pi\Delta _\infty (\hat {{\bf p}})}{\sqrt{
|\Delta _\infty (\hat {{\bf p} })|^2-\omega ^2}}\!\!\enspace ,
\end{equation}
must be also taken account of as the additional condition for the solutions
of Eqs.(\ref{ee}).

Supposing that one can choose the gap function $\Delta$ to be real within
the superconducting half space with impenetrable boundary (i.e. in the
absence of a current across the junction), we define
\begin{equation}
\label{f12}f_1=\frac 12\left( f-f^{+}\right) \enspace ,\enspace
f_2=\frac 12\left( f+f^{+}\right) \enspace .
\end{equation}

The Eqs.(\ref{ee}), (\ref{n}), being applied to $f_{1,2}$, take the form
\begin{equation}
\label{ee12}\left\{
\begin{array}{l}
2\omega f_1+iv_{x}\partial _{x}f_2+2\Delta g=0 \\
\\
f_2=-i
\frac{\displaystyle v_x}{\displaystyle 2\omega } \partial _{x}f_1 \\  \\
\partial _{x}g=-i\frac{\displaystyle 2\Delta }{\displaystyle v_x}f_2
\enspace ,
\end{array}
\right.
\end{equation}

\begin{equation}
\label{n12}g^2+f_2^2-f_1^2=-\pi^2\enspace .
\end{equation}

The boundary conditions at $x=0$ for functions $f_{1,2}$ are the same as
(\ref{bc0}), while in the depth of the superconductor we have

\begin{equation}
\label{bc012}f_1\left( \hat {{\bf p}},\omega \right) \mid _{x=\infty }=\frac{
\pi\Delta _\infty (\hat {{\bf p}})}{\sqrt{|\Delta _\infty (\hat {{\bf p}
})|^2-\omega ^2}}\enspace ,\enspace
f_2\mid _{x=\infty }=0\enspace .
\end{equation}

The representations for $g$ and $f$ taken on the boundary, which turn out to
be useful from the point of view of the consideration of singular parts of
the propagators, may be derived from these equations. For obtaining the
representations it is convenient to introduce the following function
\begin{equation}
\label{dd}\tilde \Delta (\hat {{\bf p}},\omega )=\frac{\int_0^\infty \Delta
(\hat {{\bf p}},x)f_2(\hat {{\bf p}},x,\omega )dx}{\int_0^\infty f_2(\hat {
{\bf p}},x,\omega )dx} \enspace .
\end{equation}

Since from the second and third equations of (\ref{ee12}) one easily obtains
\begin{equation}
\label{g0}g(\infty )-g(0)\!=\!-\frac{2i}{v_x}\!\int_0^\infty \!\!\!\!\Delta
(\hat {{\bf p}},x)f_2(\hat {{\bf p}},x,\omega )dx\!\enspace ,
\enspace f_1(\infty)-f_1(0)=\frac{2i\omega }{v_x}\!\int_0^\infty\!\! f_2(\hat
{{\bf p}},x,\omega )dx
\end{equation}

the quantity $\tilde \Delta (\hat {{\bf p}},\omega )$ may be written also in
the form
\begin{equation}
\label{del}\tilde \Delta (\hat {{\bf p}},\omega )=-\omega \frac{g(\infty
)-g(0)}{f_1(\infty )-f_1(0)}\enspace .
\end{equation}

Here and below we denote
$$
g\mid _{x=0}=g(0)\, ,\quad g\mid _{x=\infty }=g(\infty )\, ,\quad f\mid
_{x=0}=f(0)\, ,\quad f\mid _{x=\infty }=f(\infty )\enspace .
$$

After the substitution of the expressions (\ref{bcd}), (\ref{bc012}) for the
propagators in the depth of the superconductor into the Eq.(\ref{del}), it
reduces to the following relation between $g(0)$, $f_1(0)$ and $\tilde
\Delta (\hat {{\bf p}},\omega )$:

\begin{equation}
\label{g2}g(0)=\frac{\pi}{\omega}\frac{\tilde \Delta (\hat {{\bf p}},\omega
) \Delta _\infty(\hat {{\bf p}})-\omega ^2}{\sqrt{|\Delta _\infty (\hat {
{\bf p}})|^2-\omega ^2}}-\frac{\tilde \Delta (\hat {{\bf p}},\omega )}\omega
f_1(0)\enspace .
\end{equation}

Entirely analogous relation may be written also for the momentum direction $
\check {{\bf p}}$, after that the boundary conditions (\ref{bc0}) allow us
to write down the following representations for $g(0)\, $and $f_1(0)\, $
separately

\begin{equation}
\label{g1}g(0)\!=\!\frac \pi \omega \frac{\tilde \Delta (\hat {{\bf p}
},\omega )\tilde \Delta (\check {{\bf p}},\omega )}{\tilde \Delta (\check {
{\bf p}},\omega )\!-\!\tilde \Delta (\hat {{\bf p}},\omega )}\!\!\left[\!
\frac{ \Delta _\infty (\hat {{\bf p}})\tilde \Delta (\hat {{\bf p}},\omega
)-\omega ^2}{\tilde \Delta (\hat {{\bf p}},\omega )\sqrt{|\Delta _\infty (\hat
{{\bf p }})|^2\!-\!\omega ^2}}-\!\frac{\Delta _\infty (\check {{\bf p}})\tilde
\Delta (\check {{\bf p}},\omega )-\omega ^2}{\tilde \Delta (\check {{\bf
p}},\omega )\sqrt{|\Delta _\infty (\check {{\bf p}})|^2\!-\!\omega^2}}\!
\right]
\end{equation}

\begin{equation}
\label{f1}f_1(0)=\frac{\pi}{\tilde \Delta (\hat {{\bf p}},\omega )-\tilde
\Delta (\check {{\bf p}},\omega )}\left[ \frac{\Delta _\infty (\hat {{\bf p}
})\tilde \Delta (\hat {{\bf p}},\omega )-\omega ^2}{\sqrt{|\Delta _\infty
(\hat {{\bf p}})|^2-\omega ^2}}- \frac{\Delta _\infty (\check {{\bf p}
})\tilde \Delta (\check {{\bf p}},\omega )-\omega ^2}{\sqrt{|\Delta _\infty
(\check {{\bf p}})|^2-\omega ^2}}\right] \enspace .
\end{equation}

One can see from Eq.(\ref{g1}) that candidates for the singular points
of the propagator $g(0)$ are $\omega=0 , \, \pm |\Delta _\infty (\hat {{\bf p
}})|, \, \pm |\Delta _\infty (\check {{\bf p}})|$, and, generally speaking,
the singularities of functions $\tilde \Delta (\hat {{\bf p}},\omega ),
\, \tilde \Delta (\check {{\bf p}},\omega )$. Analogously, one finds from
Eq.(\ref{f1}) that candidates for the singular points of $f_1(0)$ are
$\omega=\pm |\Delta_\infty (\hat {{\bf p}})|, \,
\pm |\Delta _\infty (\check {{\bf p}})|$. As far as the solutions of the
equality $\tilde \Delta (\hat {{\bf p}},\omega)=
\tilde \Delta (\check {{\bf p}},\omega )$ are concerned, we note that the
consideration of Eq.(\ref{g2}) for the momentum direction $\check {{\bf p}}$
doesn't result then in the independent relation as compared to Eq.(\ref{g2})
for the momentum direction $\hat {{\bf p}}$. Thus, some additional
information is needed for the consideration of these limiting cases in
Eqs.(\ref{g1}), (\ref{f1}).

Let us consider firstly singular parts of the propagators $g_s$, $f_s$, $
f^+_s$ taken at the boundary in the vicinity of the point $\omega=0$. Taking
the low-frequency limit $\omega \rightarrow 0$ in Eqs.(\ref{g1}), (\ref{f1}),
we find that the function $f_1(0)$ \, has no singularity at $\omega=0$,
opposite to the propagator $g(0)$ which turns out to have the pole at this
point ( under the condition that $\Delta _\infty (\hat {{\bf p}})$ \, and $
\Delta _\infty (\check {{\bf p}})$\, have opposite signs):
\begin{equation}
\label{gs1}g_s(0)=\frac 1\omega \frac{\pi\tilde \Delta (\hat {{\bf p}
},0)\tilde \Delta (\check {{\bf p}},0)}{\tilde \Delta (\check {{\bf p}
},0)-\tilde \Delta (\hat {{\bf p}},0)}\left( sgn(\Delta _\infty (\hat {{\bf p
}}))-sgn(\Delta _\infty (\check {{\bf p}}))\right) =\frac{B_g(\hat {{\bf p}})}
\omega \enspace .
\end{equation}

It is remarkable that in addition to this relation one may find also the
explicit expression for the quantity $\tilde \Delta (\hat {{\bf p}},0)$
through the inhomogeneous distribution of the order parameter. Indeed, since
the function $f_1$ has no singularity at $\omega =0$ (see Eq.(\ref{ee12})),
we can derive from the normalization condition (\ref{n12}) the relation $
g_s=\pm if_{2s}$ (when $\omega\rightarrow 0$) between the singular parts of $
g$ and $f_2$. Then from the last equation of the system (\ref{ee12}) we get
the equation for the singular part of $f_2$:\, $\partial _xf_{2,s}\pm \frac{
2\Delta }{v_x}f_{2,s}=0$. The solution of this equation which satisfies the
conditions (\ref{bc012}), (\ref{bc0}), is
\begin{equation}
\label{f20}f_{2,s}(\hat {{\bf p }}, x, \omega)= f_{2,s}(\hat {
{\bf p }},0,\omega) \exp{\left( -\frac{2sgn(\Delta _\infty(\hat {{\bf p}}))}{
|v_x|}\int_0^x \Delta(\hat {{\bf p}},x^{\prime })dx^{\prime }\right)}.
\end{equation}

Substituting this solution into Eq.(\ref{dd}) we obtain

\begin{equation}
\label{do}\tilde \Delta (\hat {{\bf p}},0)=\frac 12\frac{\displaystyle
|v_x|sgn(\Delta
_\infty (\hat {{\bf p}}))}{\displaystyle
\int_0^\infty \exp \left( -\frac{2sgn(\Delta
_\infty (\hat {{\bf p}}))}{|v_x|}\int_0^x\Delta (\hat {{\bf p}},x^{\prime
})dx^{\prime }\right) dx}\enspace .
\end{equation}

Note, that the determination of sign in Eq.(\ref{f20}) allows to fix sign in
the relation between the zero frequency singular parts of the quasiclassical
propagators in the close vicinity of the boundary:

\begin{equation}
\label{fs1}f_s(\hat {{\bf p }}, x, \omega)= f^+_s(\hat {{\bf p }
}, x, \omega)=
-i\,sgn(v_x \Delta _\infty (\hat {{\bf p }}))g_s(\hat {{\bf p }}, x,
\omega) \enspace .
\end{equation}

Thus, Eqs.(\ref{gs1}), (\ref{do}) and (\ref{fs1}) provide the quite general
description for the zero frequency singular parts of the propagators taken
at the boundary. If one is interested in only the singular points of the I-V
curve, one can separate the problem of the solution of the self-consistency
equation for particular pairing potentials for other investigations. The
latter problem being an important part of the total theoretical description
of the I-V curve for the tunnel junction, is very cumbersome and obviously
includes large numerical investigations within the framework of any
microscopic model for the pairing potential.

Passing to the other candidates for the singular points $\omega=\pm |\Delta
_\infty (\hat {{\bf p}})|,$ \, $\pm |\Delta _\infty (\check {{\bf p}})|$, we
find firstly one important relationship for the function $\tilde \Delta
(\hat {{\bf p}},\omega )$ at this frequency. For this purpose the asymptotic
behavior of the function $f_2\left( \hat {{\bf p}},\omega\right)$ at $
x\rightarrow \infty$ is of interest, which follows from
Eqs.(\ref{ee12})--(\ref{bc012}):
\begin{equation}
\label{as}f_2\left( \hat {{\bf p}},x,\omega \right) \propto \exp \left( -
\frac{2\sqrt{|\Delta _\infty (\hat {{\bf p}})|^2-\omega ^2}}{|v_x|}x\right)
\enspace .
\end{equation}

One can see from Eqs.(\ref{as}), (\ref{dd}) that in the limit $|\omega
|\rightarrow |\Delta_\infty (\hat {{\bf p}})|$ the main contribution to the
integrals in (\ref{dd}) comes from the depth of the superconductor, where $
\Delta (\hat{{\bf p}},x)$ is equal to its bulk value. Then one obtains the
relationship
\begin{equation}
\label{p}\tilde \Delta (\hat {{\bf p}},\omega )\rightarrow \Delta _\infty
(\hat {{\bf p}})\enspace \quad {\it for} \qquad |\omega |\rightarrow |\Delta
_\infty (\hat {{\bf p}})|\enspace .
\end{equation}

Now one can consider the behaviors of $g$ and $f_1$ in the vicinity of the
points $|\omega |\rightarrow |\Delta_\infty (\hat {{\bf p}})|, \, |\Delta
_\infty (\check {{\bf p}})|$. Taking the limit $|\omega|\rightarrow |\Delta
_\infty (\hat {{\bf p}})|$\, {\it or} \, $|\Delta _\infty (\check {{\bf p}
})| $ in Eqs.(\ref{g1}), (\ref{f1}), one can find with the help of (\ref{p})
the cancellation of divergences and, in the first approximation, the
appearance of the square root nonanalitycal behavior of the form $\sqrt{
\Delta_\infty ^2\left( \hat {{\bf p}}\right) -\omega ^2}$ (or $\sqrt{
\Delta_\infty ^2\left( \check {{\bf p}}\right) -\omega ^2}$) for $g(0)$ and $
f_1(0)$ (and therefore $f(0)$) at these points. The exclusions may represent
the orientations for which $\tilde \Delta (\hat {{\bf p}})=\tilde\Delta
(\check {{\bf p}})$, when strictly speaking Eqs.(\ref{g1}), (\ref{f1})
themselves do not provide more information than (\ref{g2}). It is well known
that if this condition holds for all momentum orientations, there is no
surface pair breaking and the divergences in the propagators at $|\omega
|\rightarrow |\Delta_\infty (\hat {{\bf p}})|$ occur in this particular
case:
\begin{equation}
\label{gb}g^R\left( \hat{{\bf p}},\omega \right) =-\frac{\pi\omega}{\sqrt{
|\Delta (\hat{{\bf p}})|^2-\omega ^2}}\enspace ,\enspace
f^R\left( \hat{{\bf p}},\omega \right) =\frac{\pi\Delta (\hat{{\bf p}})}{
\sqrt{|\Delta ( \hat{{\bf p}})|^2-\omega ^2}}\enspace .
\end{equation}

At last one should discuss the possibility for existence of quasiparticle
bound states with nonzero energy located near the boundary. They may
appear, for instance, due to the spatial dependence of the order parameter
which is suppressed at the boundary. The bound state may be interpreted as a
bound state in the "potential well" formed by the order parameter \cite
{buch2}. Since the quasiparticle bound state corresponds to the pole in the
quasiclassical propagators, we simply add below the pole-like term to the
singular parts of the propagators. If the propagators $g(0)$, $f_1(0)$
have similar poles describing localized states with nonzero energy, then it
follows from Eq.(\ref{del}) and the boundary conditions (\ref{bc0}), that
at the point of the pole $\tilde \Delta (\hat {{\bf p}},\omega)=
\tilde \Delta (\check {{\bf p}},\omega )$.

Taking into account the results obtained above, the nonanalytical terms of $
g(0)$ and $f(0)$ may be written in the form
$$
g_s^R\left( \hat {{\bf p}},\omega \right) \mid _{x=0}=\frac{B_g\left( \hat {
{\bf p}}\right) }{\omega +i\delta }+\frac{Q_g\left( \hat {{\bf p}}\right) }{
\omega -h(\hat {{\bf p}})\cdot sgn(\omega )+i\delta }+
$$

\begin{equation}
\label{gis}+C\left( \hat {{\bf p}},\omega \right) sgn(\omega ) \sqrt{
\Delta_\infty ^2\left( \hat {{\bf p}}\right) -\omega ^2}+C\left( \check {
{\bf p}},\omega \right) sgn(\omega )\sqrt{\Delta _\infty ^2\left( \check {
{\bf p}}\right) -\omega ^2}+ \cdots \enspace ,
\end{equation}

$$
f_s^R\left(\hat{{\bf p}},\omega\right) \mid _{x=0}=\frac{ iB_f\left( \hat {
{\bf p} }\right) }{\omega +i\delta } + \frac{iQ_f\left( \hat {{\bf p}
}\right) }{\omega -h(\hat {{\bf p}})\cdot sgn(\omega )+i\delta }+
$$

\begin{equation}
\label{fis}+E\left( \hat {{\bf p}},\omega \right) \sqrt{\Delta _\infty
^2\left( \hat {{\bf p}}\right) -\omega ^2}+E\left( \check {{\bf p}},\omega
\right) \sqrt{\Delta _\infty ^2\left( \check {{\bf p}}\right) -\omega ^2}+
\cdots \enspace , \enspace \delta\rightarrow +0 \enspace .
\end{equation}
The following relationships take place: \, $B_f(\hat {{\bf p}})=
-sgn(v_x \Delta_{\infty}(\hat {{\bf p}}))B_g(\hat {{\bf p}})=
-B_f(-\hat {{\bf p}})$, $Q_g(\hat {{\bf p}})=|Q_f(\hat {{\bf p}})|$,
$Q_f(\hat {{\bf p}})=-Q_f^*(-\hat {{\bf p}})$, $h(-\hat {{\bf p}})=
h(\hat {{\bf p}})$.

Below we are interested  only in the values of functions
$Q_{g,f}(\hat {{\bf p}})$ near the poles $\omega=\pm h(\hat {{\bf p}})$.
In considering positive and negative poles one should take into account
that quantities $Q_g(\hat {{\bf p}},\omega)$, $Re Q_f(\hat {{\bf p}},\omega)$
are even functions, while $Im Q_f(\hat {{\bf p}},\omega)$ is an odd function
of $\omega$. We do not indicate further the explicit energy dependence of
functions $Q_{g,f}$, using only their values at the positive pole.

Since for $|\omega| <min\left( |\Delta _\infty (\check {{\bf p}})|, |\Delta
_\infty (\hat {{\bf p}})|\right)$ the quasiparticle density of states in the
continuum is zero for fixed momentum direction, it is natural to demand $
Im\,C, Im\,Q_{g}=0$ under this condition.
In the particular case $\Delta (\check {{\bf p}},x)=-\Delta
(\hat {{\bf p}},x)$ , when $\Delta(x=0),\, f_1(0)=0$, one may obtain from
the Eilenberger equation $Re\, f(0)=0$ for the frequencies $
|\omega|\!<|\Delta _\infty \! (\hat {{\bf p}})|$. Then for this frequency
interval one has $Re\left( E\left( \hat {{\bf p}},\omega \right) +E\left(
\check {{\bf p}}, \omega \right) \right)\!= 0$.

As we show below, the location and type of singular points on the I-V
characteristics of the tunnel current associated with the
quasiparticle bound states, are determined, in particular, by the
extremal and nonanalitical points of the function $h(\hat{{\bf p}})$.
From the boundary conditions for the propagators and the parity of
pairing we get $h(\hat{{\bf p}})=h(-\check{{\bf p}})$. Taking into account
also that for $\hat{{\bf p}}\parallel {\bf n}$ it follows $\hat{{\bf p}}=-
\check{{\bf p}}$, we find that the function $h(\hat{{\bf p}})$ (as well as
the total propagator $g(0)$) must have an extremal value for the direction
$\hat{{\bf p}}$ along the normal to the boundary ${\bf n}$.
Analogously, we get that the function
$Q_f(\hat{{\bf p}})$ takes purely imaginary value for the direction
$\hat{{\bf p}}\parallel {\bf n}$. In particular, it follows from here for the
the orientations  $\Delta (\check {{\bf p}},x)=-\Delta
(\hat {{\bf p}},x)$, that $Q_{f,g}(\hat{{\bf p}})=0$  for the momentum
direction $\hat{{\bf p}}\parallel {\bf n}$. The other characteristic points of
the function $h(\hat{{\bf p}})$ are the momentum
directions where the quasiparticle state bound to the boundary disappears.

The results of numerical calculations of quantities $B_g(\phi )$,
$Q_g(\phi )=|Q_f(\phi )|$, $ReQ_f(\phi )$ and $h(\phi )$ are presented in
Fig. 1. We consider a tetragonal superconductor with cylindrical Fermi surface
and impenetrable specularly reflecting interface
at $x=0$ (the cylindrical axis $z$ is parallel to the boundary
plane). The pairing interaction is supposed to have the particular form
$V(\phi ,\phi^{\prime })=2V\cos (2\phi -2\phi _o)\cos (2\phi ^{\prime }-
2\phi _o)$, which results in a d-wave order parameter. Here $\phi $ is the
azimuthal angle in $xy$ -- plane, which is counted from the direction of the
normal to the boundary.
Angle $\phi _o$ describes orientation of the crystalline $x_0$-axis
with respect to the normal to the boundary. Then the order parameter
has the form: $\Delta (\phi ,x)=\Delta(x)\cos (2\phi -2\phi _o)$,
where $\Delta (x)$ has to be evaluated
self-consistently. The angle $\phi $ defines the direction of incoming
momentum along the quasiparticle trajectory. We choose $\phi _o=\pi /9$,
$T=0.45T_c$, $\Delta_0/(2T)=2$,
where $\Delta _0\equiv \Delta (x=\infty )$,
$T_c$ is the critical temperature. For
the surface to lattice orientation $\phi _o=\pi /9$ the order parameter is
suppressed in the vicinity of the boundary up to the value $\Delta
(x=0)=0.28\Delta _0$. Quantities $B_g(\phi )$, $Q_g(\phi )$, $ReQ_f(\phi )$
are normalized in Fig. 1 to the value $\pi \Delta _0$, while the quantity $
h(\phi )$ -- to the value $\Delta _0$.
One can see that the bound states with
nonzero energy $\pm h(\phi )$ exist only in the narrow angular region $\phi
\in (-0.095,0.095)$ in the vicinity of the normal to the boundary.
The mid-gap states exist within two other wider regions of momentum
directions, where  $B_g(\phi )\neq 0$.
It follows from Fig. 1,  that the maximal value
of the function $h(\phi )$ is $h_m=0.7\Delta _0$ and the value of $h(\phi )$
at the edge points is $h_{ed}=0.63\Delta _0$. It is worth noting that
for the momentum directions $\hat {{\bf p}}$ ($\phi_{ed}= \pm 0.095$),
where the bound states with nonzero quasiparticle energy disappear,
the following relation holds $h_{ed}=
min\left( |\Delta _\infty (\check {{\bf p}})|,
|\Delta _\infty (\hat {{\bf p}})|\right)$. This means that
discrete bound levels combine with the continuum spectrum at
the momentum directions $\phi_{ed}$. Quantities
$C\left( \hat {{\bf p}},\omega \right)$,
$E\left( \hat {{\bf p}},\omega \right)$ (or
$C\left( \check {{\bf p}},\omega \right)$,
$E\left( \check {{\bf p}},\omega \right)$ - depending on what quantity is
smaller among $ |\Delta _\infty (\hat {{\bf p}})|,
|\Delta _\infty (\check {{\bf p}})|$ ), which are present in Eqs.(\ref{gis}),
(\ref{fis}), diverge at the momentum directions $\phi_{ed}$ and the energy
$|\omega|=h_{ed}$.

\section{Current-voltage Characteristics for the Josephson and Quasiparticle
Currents}

The peculiarities of the I-V curves are due to the singular points of
functions $I_m(V, \hat{{\bf p}}_1)$. If these functions don't depend upon
the momentum directions, their singular behavior with the variable $V$
directly describes the respective behavior of the I-V curves, according to
Eqs. (\ref{cur5}) -- (\ref{i4}). In contrast to this case, for
anisotropically paired superconductors the integration over momentum
directions greatly modifies the singularities of the current--voltage
characteristics. Then the singular points of functions $j_m(V)$ describe the
singularities on the I-V curves. The other distinctive feature of the
anisotropically paired superconductors which essentially
influences the behavior of the Josephson effect as well as the quasiparticle
current, is their sensitivity to the inhomogeneities and
interfaces. The suppression of the anisotropic
order parameter at the boundary results in the specific surface superconducting
effects. At least two important features must be taken into account in this
context. The former is associated with the possibility for the opposite
signs of the bulk order parameter taken for the directions of the incident
and reflected momenta. This results in the zero energy quasiparticle bound
states at the specular boundary. The latter appears if the additional
quasiparticle bound states localized near the barrier plane occur due to
the particular form of the spatial dependence of the order parameter (see,
\cite{buch1,buch2,mat1,h,yh,mat2,nag}).

\subsection{Crystal Orientations with no Surface Pair Breaking}

Specific features of current-voltage characteristics for the tunnel electric
current for anisotropically paired superconductors differ from the s-wave
isotropic ones even in the case of conventional boundary conditions, when
the surface doesn't suppress the superconducting order parameter. This holds
if the values of the order parameter taken for the incident and reflected
momenta are equal to each other for all momentum directions. The
consideration of this particular case is just the subject of current
section. Let us consider the behavior of functions $j_m(V)$, which determine
the current-voltage curves, for the cases when the electron propagators at
the impenetrable plane have the form (\ref{gb}) coinciding with their bulk
expressions. Substituting these propagators into Eqs.(\ref{i1})-(\ref{i4}),
we get
$$
I_1(V,\hat{{\bf p}}_1)=-\Delta _1(\hat{{\bf p}}_1)\Delta _2(\hat{{\bf p}}_2)
\int\limits_{-\infty }^\infty d\omega \tanh \left( \frac {|\omega|}{2T}\right)
\left(\frac{\Theta \left( |\Delta _1(\hat{{\bf p}} _1)|-|\omega -V|\right)
}{\sqrt{|\Delta _1(\hat{{\bf p}}_1)|^2- (\omega -V)^2 }}\times \right.
$$
\begin{equation}
\label{i1b}
\left. \times\frac{\Theta \left( |\omega |- |\Delta
_2(\hat{{\bf p }}_2)|\right) }{\sqrt{\omega ^2-|\Delta _2(\hat{{\bf
p}}_2)|^2}}+ \frac{ \Theta \left( |\omega |- |\Delta _1(\hat{{\bf
p}}_1)|\right) }{\sqrt{\omega ^2-|\Delta _1(\hat{{\bf p}}_1)|^2}}\frac{\Theta
\left( |\Delta _2(\hat{{\bf p }}_2)|-|\omega +V|\right) }{\sqrt{|\Delta
_2(\hat{{\bf p}}_2)|^2-(\omega +V)^2}}\right) \enspace ,
\end{equation}

$$
I_2(V,\hat{{\bf p}}_1)=\Delta _1(\hat{{\bf p}}_1)\Delta _2(\hat{{\bf p}}_2)
\int\limits_{-\infty }^\infty d\omega \left( \tanh \left(
\frac \omega {2T}\right) -\tanh\left( \frac{\omega +V}{2T}\right) \right)\times
$$

\begin{equation}
\label{i2b}\times\frac{sgn(\omega )sgn(\omega +V)\Theta \left( |\omega
|-|\Delta _1(\hat{{\bf p}}_1)|\right) }{\sqrt{\omega ^2-|\Delta _1(\hat{{\bf
p}}_1)|^2}}\frac{\Theta \left( |\omega +V|-|\Delta _2(\hat{{\bf p}}
_2)|\right) }{\sqrt{(\omega +V)^2-|\Delta _2(\hat{{\bf p}}_2)|^2}} \enspace
,
\end{equation}

$$
I_3(V,\hat{{\bf p}}_1)=\int\limits_{-\infty }^\infty d\omega \left( \tanh
\left( \frac{
\omega -V}{2T}\right) -\tanh \left( \frac \omega {2T}\right) \right) \frac{
|\omega ||\omega -V|\Theta \left( |\omega |-|\Delta _1(\hat{{\bf p}}
_1)|\right) }{\sqrt{\omega ^2-|\Delta _1(\hat{{\bf p}}_1)|^2}}\times
$$

\begin{equation}
\label{i3b}\times\frac{\Theta \left( |\omega -V|-|\Delta _2(\hat{{\bf p}}
_2)|\right) }{\sqrt{(\omega -V)^2-|\Delta _2(\hat{{\bf p}}_2)|^2}} \enspace,
\end{equation}

$$
I_4(V,\hat{{\bf p}}_1)=-\int\limits_{-\infty }^\infty d\omega \tanh \left(
\frac \omega
{2T}\right) (\omega -V)|\omega |\left(\frac{\Theta \left( |\Delta _1(\hat{
{\bf p}}_1)|-|\omega -V|\right) }{\sqrt{|\Delta _1(\hat{{\bf p}}
_1)|^2-(\omega -V)^2}}\times \right.
$$
\begin{equation}
\label{i4b}\left. \times\frac{\Theta \left( |\omega |-|\Delta _2(\hat{{\bf p}
}_2)|\right) }{\sqrt{\omega ^2-|\Delta _2(\hat{{\bf p}}_2)|^2}}+ \frac{
\Theta \left( |\omega |-|\Delta _1( \hat{{\bf p}}_1)|\right) }{\sqrt{\omega
^2- |\Delta _1(\hat{{\bf p}}_1)|^2}}\frac{\Theta \left( |\Delta _2(\hat{{\bf
p}}_2)|-|\omega -V|\right) }{\sqrt{|\Delta _2(\hat{{\bf p}}_2)|^2-(\omega
-V)^2}}\right) \enspace .
\end{equation}

The singularities of functions $I_m(V, \hat{{\bf p}}_1)$ appear after the
integration over the frequency only if two square roots (multiplied by each
other in the denominators of the integrands) are equal to zero
simultaneously. It is possible for certain values of frequency and voltage
and the singularities turn out to be located at momentum-dependent points $
|V|=\left|\left |\Delta_2\right|\pm \left|\Delta_1\right| \right|$. It
follows from Eqs.(\ref{i1b})-(\ref{i4b}) that the expressions for singular
parts of $I_m(V, \hat{{\bf p}}_1)$ are as follows

$$
I_1=\frac 12\sqrt{|\Delta _1\Delta _2|}sgn(\Delta _1\Delta _2) \left\{\left[
\tanh \left( \frac{|\Delta _1|}{2T}\right) +\tanh \left( \frac{|\Delta _2|}{
2T} \right) \right]\times \right.
$$

\begin{equation}
\label{i1s}\left. \times\ln\left| |V|\!-\!|\Delta _1|-\!|\Delta _2|\right|\!
-\!\pi\! \left|\tanh \!\left( \frac{|\Delta _1|}{2T}\right)\!-\tanh\!\left(
\frac{|\Delta _2|}{2T}\right) \right| \Theta \left(
|V|\!-\!\left|\left|\Delta _2\right|\!-\!|\Delta _1|\right| \right) \right\}
,
\end{equation}

$$
I_2=\frac 12\sqrt{|\Delta _1\Delta _2|}sgn(\Delta _1\Delta
_2)sgn(V)\left\{sgn\left( |\Delta _1|-|\Delta _2|\right) \left[ \tanh \left(
\frac{|\Delta _1|}{2T}\right)- \right. \right.
$$

$$
\left. -\tanh \left( \frac{|\Delta _1|+|V|sgn\left( |\Delta _2|-|\Delta
_1|\right) }{2T}\right) \right] \ln \left| |V|-\left| |\Delta _2|-|\Delta
_1|\right| \right|+
$$

\begin{equation}
\label{i2s}\left. +\pi \left[ \tanh \left( \frac{|\Delta _1|}{2T}\right)
+\tanh \left( \frac{|\Delta _2|}{2T}\right) \right] \Theta \left(
|V|-|\Delta _1|-|\Delta _2|\right) \right\} \enspace ,
\end{equation}

$$
I_3=\frac 12\sqrt{|\Delta _1\Delta _2|}sgn(V)\left\{sgn\left( |\Delta
_1|-|\Delta _2|\right) \left[ \tanh \left( \frac{|\Delta _1|}{2T}\right)-
\right. \right.
$$
$$
\left. -\tanh \left( \frac{|\Delta _1|+|V|sgn\left( |\Delta _2|-|\Delta
_1|\right) }{2T}\right) \right] \ln \left| |V|-\left| |\Delta _2|-|\Delta
_1|\right| \right| -
$$

\begin{equation}
\label{i3s}\left. -\pi \left[\tanh \left( \frac{|\Delta _1|}{2T}\right) +
\tanh \left( \frac{|\Delta _2|}{ 2T}\right) \right] \Theta \left(
|V|-|\Delta _1|-|\Delta _2|\right) \right\} \enspace ,
\end{equation}

$$
I_4=-\frac 12\sqrt{|\Delta _1\Delta _2|}\left\{ \left[ \tanh \left( \frac{
|\Delta _1|}{2T}\right) +\tanh \left( \frac{|\Delta _2|}{2T}\right) \right]
\ln \left| |V|-|\Delta _1|-|\Delta _2|\right| + \right.
$$

\begin{equation}
\label{i4s}\left. +\pi \left| \tanh \left( \frac{|\Delta _1|}{2T}\right)
-\tanh \left( \frac{|\Delta _2|}{2T}\right) \right| \Theta \left( |V|-\left|
|\Delta _2|-|\Delta _1|\right| \right) \right\} \enspace .
\end{equation}

According to Eqs.(\ref{cur5}), (\ref{jm}), for obtaining the current-voltage
relation in the vicinities of singular points one should carry out the
integration of expressions (\ref{i1s})-(\ref{i4s}) over the momentum
directions and consider the correspondent expressions for $j_m(V)$ . It
follows from this integration that nonanalytical behavior of the I-V curves
takes place only for the values $|V|$ in the close vicinities of the
extremal points of the expressions $\left| |\Delta _2(\hat{{\bf p}}_2)| \pm
|\Delta _1(\hat{{\bf p}}_1)|\right|$, which are considered, for example, as
functions of $\hat{{\bf p}}_1$. The characteristic behaviors of the I-V
curves near these extrema turn out to be strongly dependent upon the type of
the extremal point. Comparing the expressions (\ref{i1s})-(\ref{i4s}) one
finds the same singular behaviors for pairs of functions $j_1(V)$, $j_4(V)$
and $j_2(V)$ , $j_3(V)$ (disregarding the differences in signs for the
moment). Due to this fact it is sufficient to describe, for example, only
the singular points for $j_1(V)$ and $j_3(V)$.

It is convenient further to examine the singular points for the
conductance $G=dj_x/dV$. It can be shown, that only the derivatives of $
\Theta$- and logarithmic functions with respect to the voltage are
associated with nonanalytical behavior of $G(V)$. In the former case one
obtains $\delta$-function, which in fact reduces the integration over the
Fermi-surface to the integration over the line on the surface. Corresponding
terms $\tilde G_{1,3}(V)$ of functions $G_{1,3}(V)=dj_{1,3}/dV$ may be
represented as follows:

\begin{equation}
\label{tG1}\tilde G_1=-sgn(V)\!\int \!\!dl\frac{K^{-}\left( \hat {{\bf p}
}_1\right) }{\left| \nabla _{\hat {{\bf p}}_1}\left( |\Delta _1|-|\Delta
_2|\right) \right| }\enspace ,\enspace \tilde G_3=-\int dl\frac{K^{+}\left(
\hat {{\bf p}}_1\right) sgn(\Delta _1\Delta _2)}{\left| \nabla _{\hat {{\bf p
}}_1}\left( |\Delta _1|+|\Delta _2|\right) \right| }\enspace .
\end{equation}

Here $l$ is the local coordinate along the line $|V|=||\Delta _1(\hat {{\bf p
}}_1)|\pm |\Delta _2(\hat {{\bf p}}_2)||$ on the Fermi surface (sign plus
corresponds to $\tilde G_3$ and minus -- to $\tilde G_1$). The functions
$K^{\pm }$ are determined as follows:

\begin{equation}
K^{\pm }\left( \hat {{\bf p}}_1\right) =\pi \frac{\left| \tanh \left( \frac{
|\Delta _1|}{2T}\right) \pm \tanh \left( \frac{|\Delta _2|}{2T}\right)
\right| \sqrt{|\Delta _1\Delta _2|}sgn(\Delta _1\Delta _2)\frac{v_{x1}}{
v_{f1}}D}{2\left( 2\pi \right) ^3}
\end{equation}

We consider below various types of extrema and get respective behaviors of $
G_m(V)$. Let the function $\left| |\Delta _2|-|\Delta _1|\right| $ take
maximal or minimal value at the point ${\bf p}_1={\bf p}_0$ on the Fermi
surface and in the vicinity of this point has the form
\begin{equation}
\label{mmp}\left| |\Delta _2|-|\Delta _1|\right| =a\pm (b\tilde
p_1^2+c\tilde p_2^2),a,b,c>0\enspace .
\end{equation}
Here $\tilde p_1$, $\tilde p_2$ are the local orthogonal coordinates in the
vicinity of the point ${\bf p}_0$. Since the function $\left| |\Delta
_2|-|\Delta _1|\right| $ comes in the expressions for $j_{1,4}$ as the
argument of $\Theta $-function and in the formulae for $j_{2,3}$ as an
argument of the logarithmic function, we obtain two different singular
behaviors near the value $|V|=a$:
\begin{equation}
\label{dg1m}\delta G_1\mid _{|V|=a}=sgn(\Delta _1\Delta _2)\mid _{\hat {{\bf
p}}_1=\hat {{\bf p}}_0}\delta G_4\mid _{|V|=a}=\mp \frac \pi {\sqrt{bc}
}K^{-}\left( \hat {{\bf p}}_0\right)
\end{equation}

\begin{equation}
\label{g2m}G_2=sgn(\Delta _1\Delta _2)\mid _{\hat {{\bf p}}_1=\hat {{\bf p}
}_0}G_3=\pm \frac 1{\sqrt{bc}}K^{-}\left( \hat {{\bf p}}_0\right) \ln \left|
|V|-a\right|
\end{equation}
The notation for a jump of the conductance $\delta G\mid
_{|V|=a}=G(|V|>a)-G(|V|<a)$ is introduced here.

The singularities coming from the maximal and minimal values of the quantity
$|\Delta _2|+|\Delta _1|$, when one has $|\Delta _2|+|\Delta _1|=a\pm
(b\tilde p_1^2+c\tilde p_2^2),a,b,c>0$, are described analogously:

\begin{equation}
\label{dg2m}\delta G_2\mid _{|V|=a}=-sgn(\Delta _1\Delta _2)\mid _{\hat {
{\bf p}}_1=\hat {{\bf p}}_0}\delta G_3\mid _{|V|=a}=\mp \frac \pi {\sqrt{bc}
}K^{+}\left( \hat {{\bf p}}_0\right)
\end{equation}

\begin{equation}
\label{g1m}G_1=-sgn(\Delta _1\Delta _2)\mid _{_{\hat {{\bf p}}_1=\hat {{\bf p
}}_0}}G_4=\pm \frac 1{\sqrt{bc}}K^{+}\left( \hat {{\bf p}}_0\right)
sgn(V)\ln \left| |V|-a\right|
\end{equation}

In the case of a saddle point of the function $\left| |\Delta _2|-|\Delta
_1|\right| $ one has near this point
\begin{equation}
\label{sad1}\left| |\Delta _2|-|\Delta _1|\right| =a+b\tilde p_1^2-c\tilde
p_2^2, \enspace a,b,c>0.
\end{equation}
The respective singular parts of the conductance read

\begin{equation}
\label{g1s-}G_1=sgn(\Delta _1\Delta _2)\mid _{\hat {{\bf p}}_1=\hat {{\bf p}
}_0}G_4=\frac 1{\sqrt{bc}}K^{-}\left( \hat {{\bf p}}_0\right) sgn(V)\ln
\left| |V|-a\right|
\end{equation}

\begin{equation}
\label{g2s-}G_2=sgn(\Delta _1\Delta _2)\mid _{\hat {{\bf p}}_1=\hat {{\bf p}
}_0}G_3=\frac 2{\pi \sqrt{bc}}sgn(|V|-a)K^{-}\left( \hat {{\bf p}}_0\right)
\ln ^2\left| |V|-a\right|
\end{equation}

Analogously, for the saddle point
\begin{equation}
|\Delta _2|+|\Delta _1|=a+b\tilde p_1^2-c\tilde p_2^2, \enspace a,b,c>0
\end{equation}
we get

\begin{equation}
\label{g2s+}G_2=-sgn(\Delta _1\Delta _2)\mid _{\hat {{\bf p}}_1=\hat {{\bf p}
}_0}G_3=-\frac 1{\sqrt{bc}}K^{+}\left( \hat {{\bf p}}_0\right) \ln \left|
a-|V|\right|
\end{equation}

\begin{equation}
\label{g1s+}G_1=-sgn(\Delta _1\Delta _2)\mid _{\hat {{\bf p}}_1=\hat {{\bf p}
}_0}G_4=\frac{2sgn(V)sgn(|V|-a)}{\pi \sqrt{bc}}K^{+}\left( \hat {{\bf p}
}_0\right) \ln ^2\left| |V|-a\right|
\end{equation}

The jumps in conductance described by Eqs.(\ref{dg1m}),(\ref{dg2m})
correspond, of course, to the kinks on the I-V curves. Logarithmic
divergences in conductance described by Eqs.(\ref{g2m}),(\ref{g1m}),(\ref
{g1s-}) and (\ref{g2s+}), result in step-like points on the I-V curves (note
that on both sides of these singular points G has the same sign). At last
the terms containing logarithm squared in Eqs.(\ref{g2s-}), (\ref{g1s+})
describe the cusps (beak-like points) on the current-voltage
characteristics. They appear in the case of saddle points of the functions $
\left| |\Delta _2|\pm |\Delta _1|\right| $ after the integration of
logarithmic singularities for $I_m$.

Let now the quantities $\left| |\Delta _2|\pm |\Delta _1|\right|$ have the
extremal values on some line $\tilde l$ on the Fermi-surface, rather than on
isolated points as it was suggested above. Then, for example, in the
vicinities of maximal or minimal values of $\left| |\Delta _2|- |\Delta
_1|\right|$ one has
\begin{equation}
\label{l-}\left| |\Delta _2|-|\Delta _1|\right| =a\pm b\tilde p_1^2 \enspace
,\qquad a,b>0 \enspace ,
\end{equation}
where $\tilde p_1$ is the local coordinate on the Fermi-surface orthogonal
to the extremal line $\tilde l$.

In this case the inverse square root singularities in the conductance appear
from one side of the voltage value $|V|=a$:
\begin{equation}
\label{g1l-}G_1=sgn(\Delta _1\Delta _2)\mid _{\tilde l}G_4=-\frac{sgn(V)}{
\sqrt{\left| |V|-a\right| }}\Theta \left( \pm \left( |V|-a\right) \right)
\!\!\int\limits_{\stackrel{\left| |\Delta _2|-|\Delta _1|\right| =a}{v_{x1}>0
} }\!\!d\tilde l\ \frac{K^{-}}{2\sqrt{b}}\enspace ,
\end{equation}

\begin{equation}
\label{g2l-}G_2=sgn(\Delta _1\Delta _2)\mid _{\tilde l}G_3=\mp \frac{\Theta
\left( \mp \left( |V|-a\right) \right) }{\sqrt{\left| |V|-a\right| }}
\int\limits_{\stackrel{\left| |\Delta _2|-|\Delta _1|\right| =a}{v_{x1}>0}
}d\tilde l\ \frac{K^{-}}{\sqrt{b}}\enspace .
\end{equation}

These singularities correspond to the vertical slope of the I-V curve from
one side of the voltage value $|V|=a$ for each of four terms presented in
the total expression for the tunnel current. For example, in the case of
maximum on the line $\tilde l$ there are the vertical slope of the curves $
j_1(V)$ and $j_4(V)$ at $|V|=a$ from side $|V|<a$, and of the curves $j_2(V)$
, $j_3(V)$ from side $|V|>a$.

Analogously, in the case of maximal or minimal value of the quantity
$|\Delta _1|+ |\Delta _2|$ on the line $\tilde l$, when near this line one
has
\begin{equation}
\label{l+}|\Delta _1|+|\Delta _2|=a\pm b\tilde p_1^2 \enspace , \qquad a,b>0
\enspace ,
\end{equation}

the singular behavior of the conductance is described as follows:

\begin{equation}
\label{g2l+}G_2=-sgn(\Delta _1\Delta _2)\mid _{\tilde l}G_3=\frac{\Theta
\left( \pm \left( |V|-a\right) \right) }{\sqrt{\left| |V|-a\right| }}
\int\limits_{\stackrel{|\Delta _1|+|\Delta _2|=a}{v_{x1}>0}}d\tilde l\ \frac{
K^{+}}{2\sqrt{b}}
\end{equation}

\begin{equation}
\label{g1l+}G_1=-sgn(\Delta _1\Delta _2)\mid _{\tilde l}G_4=\mp \frac{sgn(V)
}{\sqrt{\left| |V|-a\right| }}\Theta \left( \mp \left( |V|-a\right) \right)
\int\limits_{\stackrel{|\Delta _1|+|\Delta _2|=a}{v_{x1}>0}}d\tilde l\ \frac{
K^{+}}{\sqrt{b}}
\end{equation}

We have considered above different superconductors or at least different
crystal orientations from both sides of the junction, when not only sum $
|\Delta _1|+| \Delta _2|$ may have extremal points or lines on the Fermi
surface, but the difference $|\Delta _1|-|\Delta _2|$ also depend upon the
momentum direction and may have extrema. As a result one can't take directly
the limit $\Delta _1=\pm \Delta _2=\Delta$ in the expressions written above
for singularities, which are associated with extrema of the difference $
|\Delta _1|-|\Delta _2|$. For this particular case quantities $G_{1,4}$ have
not singular point at $V=|\Delta _1|-|\Delta _2|=0$, while for $G_{2,3}$ in
the vicinity $V=0$ ( more exactly, for $|V|\ll T$) one gets instead of
Eqs. (\ref{g2m}),(\ref{g2s-}),(\ref{g2l-})

\begin{equation}
\label{g2=}
G_2=\pm G_3=\pm \frac{\ln |V|}{2T}\int\limits_{v_{x1}>0}\frac{
d^2S_1}{\left( 2\pi \right) ^3}\frac{v_{x1}}{v_{f1}}D\left(\hat{{\bf p}}_1
\right)\frac{|\Delta |}{\cosh ^2\frac{|\Delta |}{2T}} \enspace .
\end{equation}

For isotropic $s$-wave superconductors this term is exponentially small at
law temperatures $T\ll \Delta$. In contrast to this case, for
anisotropically paired superconductors the expression in Eq.(\ref{g2=})
manifests power law temperature behavior. For instance, for the line of
nodes of the order parameter ( when in its vicinity
$|\Delta(\hat{{\bf p}})|=b|{\tilde p}_1|$ ) it follows from
Eq.(\ref{g2=}) at low temperatures $G_{2,3}\propto T\ln |V|$.

The results of numerical calculations for $j_m(v)$ ($v=V/\Delta_o$) for the
case when there is no surface pair breaking from both sides of the tunnel
barrier are shown in Figs. 2, 3. In Figure 2 the junction between
isotropically and anisotropically paired superconductors is considered: $
\Delta_1=\Delta_0 \cos(2\phi)$ , $\Delta_2=\Delta_0/2=const$. Here $\phi$ --
is the azimuth angle in $xy$ -- plane of a tetragonal superconductor (axis z
is parallel to the boundary plane). The Fermi surface is assumed to be
cylindrical for anisotropically paired superconductor (with d-wave pairing).
We let for the barrier transparency $D\propto \cos^2(\phi)$ and $
\Delta_0/(2T)=0.5$. In this case the singular points on the I-V curves are
only the points of maximum values of the quantities $||\Delta_1(\phi)|\pm|
\Delta_2||$, as the minimum corresponds to the zero value of $\Delta_1$.
Note that for $\Delta_2=2\Delta_0$ and for the same $\Delta_1(\phi)$ only
minima of the quantities $||\Delta_1(\phi)|-|\Delta_2||$ would be of
importance. In Figure 3 the case of two identical anisotropically paired
superconductors is described: $\Delta_1=\Delta_2=\Delta_0 \cos(2\phi)$. All
functions $j_m(v)$ are normalized to the value $|j_1(0)|$.

\subsection{ Crystal Orientations with Surface Pair Breaking from One Side of
the Barrier Plane}

Let us discuss now the tunnel junction between two anisotropically paired
superconductors, considering a gradual change of the crystalline orientation
of one of them relative to the barrier plane and retaining the condition $
\Delta_2(\hat{{\bf p}})=\Delta_2(\check{{\bf p}})$ to be fulfilled only for
the second superconductor. According to Eqs.(\ref{gis}), (\ref{fis}), there
are no square root divergences of the propagators for the first
superconductor, taken on the barrier plane for an intermediate crystal
orientation. Hence, the singularities, which were found in the previous
subsection, must begin to become smooth and subsequently, for large enough
deviations from the initial orientation, will disappear. At the same time,
as it was already mentioned above, some new characteristic singular
points on the I-V curves appear in this case. One kind of them turns out to
be associated with the existence of regions on the Fermi surface with the
opposite signs of the order parameter $\Delta_{1, \infty}(\hat{{\bf p}})$.
Then the zero energy quasiparticle bound state occurs at the boundary plane.
Other singularities appear if the additional
quasiparticle bound states localized near the barrier plane appear due to
the particular form of the spatial dependence of the order parameter \cite
{buch2}. As it is seen from Eqs.(\ref{gis}), (\ref{fis}), the terms in
propagators containing the factor $1/\omega $ are of importance for the
former case. For the latter case the
contribution to the current from additional poles of the propagators has to
be taken into account. So, we let the singular parts of the propagators for
the first superconductor be represented by Eqs.(\ref{gis}), (\ref{fis}),
and for the second superconductor we use Eq.(\ref{gb}).

It is very essential that in the case when zero energy bound state takes place
only from one side of the junction (while from the other side there is no
surface pair breaking), the respective singular contribution comes only to the
quasiparticle current, not to the Josephson effect. It can be shown under
the same conditions as are supposed to be fulfilled above. The point is that
for the singular parts of the propagators, associated with the mid-gap state,
one has the following property $f_s(\hat{{\bf p}})= f^+_s(\hat{{\bf p}})$.
Opposite to this equality, the relation $f(\hat{{\bf p}})= -f^+(\hat{
{\bf p}})$ holds for the orientations when there is no surface pair
breaking. Due to this reason the corresponding singular parts of $j_1$, $j_2$
reduce to zero. It is not the case for the bound states with nonzero energy,
since the function $Q_f(\hat{{\bf p}})$ is complex in contrast to the real
function $B_f(\hat{{\bf p}})$ ($B_f^+=B_f^*=B_f$, $Q_f^+=Q_f^*$).

Substituting the expressions (\ref{gis}), (\ref{fis}),
(\ref{gb}) into Eqs.(\ref{i3}), (\ref{i4}), we get for $I_{3,4}$ the
singular points which appear after the integration over $\omega$:

$$
I_3(V,\hat {{\bf p}}_1)=-B_{g1}(\hat {{\bf p}}_1)\tanh \left( \frac
V{2T}\right) \frac{|V|\Theta \left( |V|-|\Delta _2(\hat {{\bf p}}_2)|\right)
}{\sqrt{V^2-|\Delta _2(\hat {{\bf p}}_2)|^2}}-
$$
$$
-\frac{Q_{g1}(\hat {{\bf p}}_1)\sqrt{|\Delta _2|}sgn(V)}{\sqrt{2}}\left\{
\left( \tanh \left( \frac{h_1}{2T}\right) +\tanh \left( \frac{|\Delta _2|}{
2T }\right) \!\right) \frac{\Theta \left( |V|-h_1-|\Delta _2|\right) }{\sqrt{
|V|-h_1-|\Delta _2|}}+\right.
$$

\begin{equation}
\label{i1ns}\left. +\left| \tanh \left( \frac{|\Delta _2|}{2T}\right) -\tanh
\left( \frac{h_1}{2T}\right) \right| \frac{\Theta \left( \left( |V|-\left|
|\Delta _2|-h_1\right| \right) sgn\left( |\Delta _2|-h_1\right) \right) }{
\sqrt{\left( |V|-\left| |\Delta _2|-h_1\right| \right) sgn\left( |\Delta
_2|-h_1\right) }}\right\} \enspace ,
\end{equation}

$$
I_4(V,\hat {{\bf p}}_1)=B_{g1}(\hat {{\bf p}}_1)\tanh \left( \frac{|V|}{2T}
\right) \frac{|V|\Theta \left( |\Delta _2(\hat {{\bf p}}_2)|-|V|\right) }{
\sqrt{|\Delta _2(\hat {{\bf p}}_2)|^2-V^2}}+
$$
$$
+\frac{Q_{g1}(\hat {{\bf p}}_1)\sqrt{|\Delta _2|}}{\sqrt{2}}\left\{ \left(
\tanh \left( \frac{h_1}{2T}\right) +\tanh \left( \frac{|\Delta _2|}{2T}
\right) \right) \frac{\Theta \left( h_1+|\Delta _2|-|V|\right) }{\sqrt{
h_1+|\Delta _2|-|V|}}+\right.
$$

\begin{equation}
\label{i2ns}\left. +\left( \tanh \left( \frac{|\Delta _2|}{2T}\right) -\tanh
\left( \frac{h_1}{2T}\right) \right) \frac{\Theta \left( \left( |V|-\left|
|\Delta _2|-h_1\right| \right) sgn\left( h_1-|\Delta _2|\right) \right) }{
\sqrt{\left( |V|-\left| |\Delta _2|-h_1\right| \right) sgn\left( h_1-|\Delta
_2|\right) }}\right\} \enspace .
\end{equation}

The nonanalytical terms of square root type presented in Eqs.(\ref{gis}),
(\ref{fis}), are omitted in Eqs.(\ref{i1ns}), (\ref{i2ns}), since in this
case they result in jumps or divergences only for the derivatives of the
conductance, not for the current or for the conductance itself.

Further integration over the Fermi surface may lead to different kinds of
the singular points on the I-V curve, since there are various
possibilities for the behaviors of the order parameter $\Delta_2$ and
function $h_1$ with the momentum and types of the corresponding extrema on
the Fermi surface. We consider firstly the singular points of the I-V curves
coming from the terms of $1/\omega$-form in the expressions for the
propagators of the first superconductor and confine ourselves by two
examples. In the particular case of isotropic $s$-wave second superconductor
there are the inverse square root singular point on the current-voltage
characteristic:
\begin{equation}
\label{j1sr}j_3=- \tanh\!\left( \!\frac{V}{2T}\right) \frac{|V|\Theta
\left( |V|-|\Delta _2|\right) }{\sqrt{V^2-|\Delta _2|^2}}\int
\limits_{v_{x1}>0}\frac{d^2S_1}{(2\pi )^3}\frac{v_{x1}}{v_{f1}}DB_{g1}(\hat{
{\bf p}}_1) \enspace ,
\end{equation}

\begin{equation}
\label{j2sr}j_4= \tanh \left( \!\frac{|V|}{2T}\right) \frac{|V|\Theta
\left( |\Delta _2|-|V|\right) }{\sqrt{|\Delta _2|^2-V^2}}\int
\limits_{v_{x1}>0}\frac{d^2S_1}{(2\pi )^3}\frac{v_{x1}}{v_{f1}}DB_{g1}(\hat{
{\bf p}}_1) \enspace .
\end{equation}
So, the quantity $j_{3}$ diverges at $|V|=\Delta_2$ from the side $
|V|>\Delta_2$, and the quantity $j_{4}$-- from the side $|V|<\Delta_2$.
In the case $\Delta_2=0$ (S-N junction) we get from Eq.(\ref{j1sr})
that the conductance $G_3\propto \left(T\cosh^2(V/2T)\right)^{-1}$.
Then for small voltage $|V|\ll T$ it follows $G_3\propto 1/T$ and
we obtain the zero-bias anomaly of the conductance at low enough
temperatures (see also \cite{tan}).
The divergence of $G_3$ in the zero-temperature limit takes place only in the
idealized system without taking account of the factors, which broaden
delta-peaks in the quasiparticle density of states.

Further, if the anisotropic order parameter $\Delta _2$ has an extremal line
on the Fermi surface and in its vicinity one gets
\begin{equation}
|\Delta _2|=a\pm b\tilde p_1^2,\qquad a,b>0\enspace,
\end{equation}
then the kind of the singular point depends on the behavior of $B_{g1}({\bf p
}_1)$ near this line. If this function doesn't vanish in the vicinity of the
extremal line, the logarithmic divergences take place for $j_3$ for the line
of maxima and for $j_4$ for the line of minima:
\begin{equation}
j_3,j_4\propto \sqrt{\frac ab}\tanh \left( \frac a{2T}\right) \ln \left|
|V|-a\right| \enspace .
\end{equation}

In the case of vanishing of the function $B_{g1}({\bf p}_1)$ on the line,
when in its vicinity this function has the form
\begin{equation}
B_{g1}(\tilde p_1)=\beta |\tilde p_1|\enspace ,
\end{equation}
one-sided vertical slope for functions $j_{3,4}$ appears at $|V|=a$:
\begin{equation}
G_3\propto \frac{\sqrt{a}}b\tanh \left( \frac a{2T}\right) \frac{\Theta
\left( |V|-a\right) }{\sqrt{|V|-a}}\enspace ,\enspace G_4\propto \frac{\sqrt{
a}}b\tanh \left( \frac a{2T}\right) \frac{\Theta \left( a-|V|\right) }{\sqrt{
a-|V|}}\enspace .
\end{equation}

Now let us examine the singular points coming from the terms describing the
additional pole in the propagators for the first superconductor. In contrast to
the contribution from the midgap states, even in the case of surface pair
breaking from one side of the barrier plane the quasiparticle states bound to
the tunnel barier with nonzero energy contribute  not only to the
quasiparticle current but also to the Josephson effect. It takes place due to
the nonzero values of $Im Q_f(\hat{{\bf p}})$, while
$Im B_f(\hat{{\bf p}})=0$. Positions of respective singular points on the I-V
curves turn out to be associated with the extremal points on the Fermi surface
of quantities $h_1\pm |\Delta _2|$. Considering, for example, the line of
extrema $l$ for the quantity
\begin{equation}
h_1+|\Delta _2|=a\pm b\tilde p_1^2,\qquad a,b>0\enspace,
\label{hdex}
\end{equation}
we find the logarithmic divergences and jumps for $j_m$

\begin{equation}
j_1 , -j_4=M_{f,g}^{+}\left\{
\begin{array}{c}
\ln \left| |V|-a\right|  \\
\pi \Theta (|V|-a)
\end{array}
\right. \enspace ,\enspace
-j_2 , j_3\!=M_{f,g}^{+}sgn(V)\left\{
\begin{array}{c}
\pi \Theta (a-|V|) \\
\ln \left| |V|-a\right|
\end{array}
\right. \enspace .
\end{equation}

Here the upper (down) line in all four expressions corresponds to the upper
(down) sign in Eq.(\ref{hdex}). We introduce the notations:
\begin{equation}
M_g^{\pm }=\frac 1{16\sqrt{2}\pi ^3}\int\limits_{v_{x1}>0}dl\frac{v_{x1}}{
v_{f1}}D\sqrt{\frac{|\Delta _2|}b}Q_{g1}(\hat {{\bf p}}_1)\left| \tanh
\left( \frac{h_1}{2T}\right) \pm \tanh \left( \frac{|\Delta _2|}{2T}\right)
\right| \enspace ,
\end{equation}

\begin{equation}
M_f^{\pm }\!=\!\!\!\!\!\int\limits_{v_{x1}>0}\!\!\!\!dl\frac{v_{x1}}{
v_{f1}}D\sqrt{\frac{|\Delta _2|}b}\frac{sgn(\Delta_2)}{16\sqrt{2}\pi ^3}
ImQ_{f1}(\hat {{\bf p} }_1)\!\left| \tanh \left(\!
\frac{h_1}{2T}\!\right)\! \pm\! \tanh \left(\! \frac{|\Delta
_2|}{2T}\!\right)\! \right|
\!\!\enspace .
\end{equation}

If the function $\left| |\Delta _2|-h_1\right| $ has the line of extrema
\begin{equation}
\left| |\Delta _2|-h_1\right| =a\pm b\tilde p_1^2,\qquad a,b>0\enspace,
\end{equation}
then one gets

\begin{equation}
j_{1,4}=M_{f,g}^{-}\left\{ \mp \Theta \left( \pm \left( |\Delta _2|-h_1\right)
\right) \ln \left| |V|-a\right| +\pi \Theta (\mp (|\Delta _2|-h_1))\Theta
(a-|V|)\right\}\! \enspace ,
\end{equation}

\begin{equation}
j_{2,3}=M_{f,g}^{-}sgn(V)\!\left\{\! \Theta\! \left( \mp \left(
|\Delta _2|\!-\!h_1\right) \right) \ln \left| |V|\!-\!a\right|\!  \pm\! \pi
\Theta\!  (\pm (|\Delta _2|\!-\!h_1))\Theta (a\!-\!|V|)\!\right\}\!  .
\end{equation}

So, for positive (negative) value of $|\Delta _2|-h_1$ the function $j_{2,3}$
($j_{1,4}$) has logarithmic divergence only for the line of maxima of $
\left| |\Delta _2|-h_1\right| $, while $j_{1,4}$ ($j_{2,3}$) has such
divergence only for the line of minima of the same quantity.

As it was mentioned above, the function $h(\hat{{\bf p}})$ may manifest the
nonanalytical behavior, for example, at the momentum direction, for which
the bound state at the boundary disappears. Taking this possibility into
account, we suppose now that the function $h_1+|\Delta _2|$ has near some
line on the Fermi surface the following nonanlytical form

\begin{equation}
\label{lb}h_1+|\Delta _2|=a+\left(b\Theta (\tilde p)+c\Theta (-\tilde
p)\right)\tilde p \enspace ,a>0\enspace .
\end{equation}
Particular cases $b$ (or $c$) $\rightarrow\infty$ just correspond to the
absence of the bound state for $\tilde p>0$ (or $\tilde p<0$) on this line.
Then one-sided vertical slope for functions $j_m$ appears at $|V|=a$:

\begin{equation}
G_1,-G_4=\frac{\Theta \left( a-|V|\right) }{\sqrt{a-|V|}}
sgn(V)P_{f,g}^{+}\enspace ,\enspace G_2,-G_3=\frac{\Theta \left(
|V|-a\right) }{\sqrt{|V|-a}}P_{f,g}^{+}\enspace .
\end{equation}

We introduce the notations:

\begin{equation}
P_g^{\pm }=\!\!\!\!\int\limits_{v_{x1}>0}\!\!\!dl\left( \frac
1b-\frac 1c\right) \frac{v_{x1}}{v_{f1}}D\sqrt{|\Delta _2|}
\frac{Q_{g1}(\hat {{\bf p}}_1)}{8\sqrt{2}\pi ^3} \left| \tanh \left(
\frac{h_1}{2T}\right) \pm \tanh \left( \frac{ |\Delta _2|}{2T}\right) \right|
 \enspace ,
\end{equation}

\begin{equation}
P_f^{\pm }\!=\!\!\!\!\!\int\limits_{v_{x1}>0}\!\!\!\!\!dl\left(\! \frac
1b\!-\!\frac 1c\!\right)\! \frac{v_{x1}}{v_{f1}}D\sqrt{|\Delta _2|}
\frac {sgn(\Delta_2)}{8\sqrt{2}\pi ^3}\!
ImQ_{f1}\left(\hat{{\bf p}}_1\right)\left| \tanh\! \left(\! \frac{h_1}{2T}
\!\right)\! \pm \!\tanh \!\left(\! \frac{|\Delta _2|}{2T}\!\right)\! \right|
\end{equation}

If the function $\left| h_1-|\Delta _2|\right| $ has the form (\ref{lb})
near some line, then also one-sided vertical slope for functions $j_m$
appears at $|V|=a$:

\begin{equation}
G_{1,4}=sgn(V)sgn\left( |\Delta _2|\!-\!h_1\right)
\frac{\Theta
\left( \left( a-|V|\right) \left( |\Delta _2|-h_1\right) \right) }{\sqrt{
\left( a-|V|\right) sgn\left( |\Delta _2|-h_1\right) }}
P_{f,g}^{-}
\enspace ,
\end{equation}

\begin{equation}
 G_{2,3}=-\frac{\Theta \left( \left(
|V|-a\right) \left( |\Delta _2|-h_1\right) \right) }{\sqrt{\left(
|V|-a\right) sgn\left( |\Delta _2|-h_1\right) }}P_{f,g}^{-}
\enspace  .
\end{equation}

The Eqs.(\ref{i1ns}), (\ref{i2ns}) are not appropriate for direct consideration
of the case $\Delta_2=0$ for a finite voltage, since the voltage is supposed to
lie near the corresponding singularity for each pole-like term. For S-N junction
it follows from Eqs. (\ref{jm}), (\ref{i3}), (\ref{i4}), that in the case
of extremal point of the quantity $h_1$ ($h_1=a\pm b\tilde p_1^2,\, a,b>0$ )
there is the low-temperature anomaly in the conductance for $|V|=a$:
$G_{3,4}\propto 1/\sqrt{bT}$.

The results of numerical calculations for $j_m(v)$ ($v=V/\Delta _0$) for the
case when there is surface pair breaking from one side of the tunnel barrier
are shown in Figs. 4.1, 4.2. The tunnel junction between $d$-wave and
isotropic $s$-wave superconductors is considered under the conditions:
$\Delta _{1\infty}=\Delta _0\cos (2\phi -2\phi _o)$, $\Delta _2=
0.2\Delta _0=const$. For the
$d$-wave superconductor we choose the same parameters as earlier (see Fig. 1):
$\phi_o=\pi /9$, $T=0.45T_{c1}$, $\Delta _0/(2T)=2$. Here $T_{c1}$ is the
critical temperature for the superconductor with d-wave pairing. We let for
the barrier transparency $D\propto \cos ^2(\phi )$.

There are the inverse square root singular points on the curves $j_3$, $j_4$
 for $V=\Delta _2=0.2\Delta _0$. For $V=h_m-\Delta _2=0.5\Delta _0$
there are logarithmic divergences of $j_1$, $j_4$ and jumps of $j_2$, $j_3$.
For $V=h_m+\Delta _2=0.9\Delta _0$ there are logarithmic
divergences of $j_2$, $j_3$ and jumps of $j_1$, $j_4$. At voltages
$V=h_{ed}\pm \Delta _2=0.83\Delta _0$, $0.43\Delta _0$ there are kinks on the
I-V curves, though some of them are feebly marked. All
functions $j_m(v)$ are normalized to the value $j_1(0)$.

\subsection{Crystal Orientations with Surface Pair Breaking for both
Superconductors}

Let now the both superconductors have the intermediate orientations
relative to the boundary. Then one should make use the expressions (\ref{gis}),
(\ref{fis}) for the singular parts of the propagators on both banks of the
barrier plane. Substituting these expressions into Eqs.(\ref{i1})--(\ref{i4}),
we obtain several kinds of the singularities. They come from the
multiplications of the terms with poles by each other and from the products
of the pole-like term and the square root nonanalytical term, represented in
Eqs.(\ref{gis}), (\ref{fis}). All these terms result, for example, in the
following singular part of $I_3$
$$
I_3=\frac 1\pi \tanh \!\left( \!\frac V{2T}\!\right) \!B_{g1}\left( \hat {
{\bf p}}_1\right) \!\Bigl(\!\!ReC_2\left( \hat {{\bf p}}_2\right) \sqrt{
V^2-\Delta _{2\infty }^2\left( \hat {{\bf p}}_2\!\right) }\Theta \!\left(
\!|V|-|\Delta _{2\infty }\left( \hat {{\bf p}}_2\right) |\right) \!+
$$
$$
ImC_2\left( \hat {{\bf p}}_2\right) \sqrt{\Delta _{2\infty }^2\left( \hat {
{\bf p}}_2\right) -V^2}\Theta \left( |\Delta _{2\infty }\left( \hat {{\bf p}
}_2\right) |\!-\!|V|\right) -\frac \pi 2Q_{g2}\left( \hat {{\bf p}}_2\right)
\delta \left( |V|-h_2\left( \hat {{\bf p}}_2\right) \right) \!+
$$
$$
+\left( \hat {{\bf p}}_2\rightarrow \check {{\bf p}}_2\right) \Bigr)+\frac
1\pi \biggl\{ \left( \tanh \left( \frac{h_1\left( \hat {{\bf p}}_1\right)
+V}{2T}\right) -\tanh \left( \frac{h_1\left( \hat {{\bf p}}_1\right) }{2T}
\right) \right) Q_{g1}\left( \hat {{\bf p}}_1\right) \cdot
$$
$$
\Bigl(-\frac{\pi}{4}Q_{g2}\left( \hat {{\bf p}}_2\right) \delta \left(
|h_1\left( \hat { {\bf p}}_1\right)\! +V\!|\!-\!h_2\left( \hat {{\bf
p}}_2\right) \right) +ReC_2\left( \hat {{\bf p}}_2\right) \sqrt{\left(
h_1\left( \hat {{\bf p}}_1\right) +V\right) ^2\!-\!\Delta _{2\infty }^2\left(
\hat {{\bf p}}_2\right) }\cdot
$$
$$
\Theta \left( \left| h_1\left( \hat {{\bf p}}_1\right) +V\right| -\left|
\Delta _{2\infty }\left( \hat {{\bf p}}_2\right) \right| \right)
+ImC_2\left( \hat {{\bf p}}_2\right) \sqrt{\Delta _{2\infty }^2\left( \hat {
{\bf p}}_2\right) -\left( h_1\left( \hat {{\bf p}}_1\right) +V\right) ^2}
\cdot
$$
\begin{equation}
\Theta \left( \left| \Delta _{2\infty }\left( \hat {{\bf p}}_2\right)
\right| -\left| h_1\left( \hat {{\bf p}}_1\right) +V\right| \right)
-(V\rightarrow -V)\Bigr)+\left( \hat {{\bf p}}_2\rightarrow \check {{\bf p}
}_2\right) \biggr\}+\left( 1\leftrightarrow 2\right) \enspace ,
\end{equation}
Analogues singularities appear in the expressions for $I_{1,2,4}$.

Here we have omitted the joint contribution from the zero frequency
singularities of the propagators from both sides of the junction. For finite
voltages this contribution is equal to zero  in the idealized system in
question, though it may be of importance for low voltages less or
comparable with the width of the broaden delta-peak in the density of states.
For dc Josephson effect this term leads to the low temperature anomaly in the
critical current and to the possibility of $0-\pi$ phase transition
\cite{bbr,tan2}.

Further integration over the momentum direction is performing below for the
particular cases of the lines of extrema for the order parameter $|\Delta
_{2\infty }\left( \hat{{\bf p}}_2\right) |$ or for quantities $\left| h_1\pm
|\Delta _{2\infty}|\right|$, $h_2$, $|h_1\pm h_2|$. Let , for example, the
order parameter $|\Delta _{2\infty }\left( \hat{{\bf p}}_2\right) |$ or the
quantity $\left| h_1\pm |\Delta _{2\infty}|\right|$ have the line of extrema
of the form
\begin{equation}
\label{1}|\Delta _{2\infty }|\enspace ,\enspace \left| h_1\pm |\Delta
_{2\infty }|\right| =a\pm b\tilde p^2\enspace ,\enspace a,b>0\enspace ,
\end{equation}
and the quantity $B_1(\hat {{\bf p}}_1)$, or $|\Delta _{2\infty }|,Q_1(\hat {
{\bf p}}_1)$ is not equal to zero on this line. Then the terms in the
conductance $G_{1,2,4}$ acquire the following logarithmic singularities
\begin{equation}
\label{log}sgn(V)G_1, G_2, sgn(V)G_4\propto \sqrt{\frac ab}\ln \left|
|V|-a\right| \enspace .
\end{equation}

At the same time the logarithmic divergence (\ref{log}) of $G_3$ appears
only in the case of the line of maxima, due to the relation $Im\, C=0$ under
the condition $|\omega| <min\left( |\Delta _\infty (\check {{\bf p}})|,
|\Delta _\infty (\hat {{\bf p}})|\right)$. For the particular orientation,
when one has $\Delta _{2\infty }(\check{{\bf p}}_2)=-\Delta _{2\infty }(\hat{
{\bf p}}_2)$, the logarithmic divergence of $G_4$ remains only for the line
of minima. Note, that the square root behavior of the propagators of the
form $\sqrt{ \Delta_\infty ^2\left( \hat {{\bf p}}\right) -\omega ^2}$
result here in the logarithmic divergences of the conductance. For some
other cases these terms may result in the kinks for a conductance or in the
divergences of its derivatives. We do not consider these kinds of
singularities here.

If there is the line of extrema of $h_2$
\begin{equation}
h_2=a\pm b\tilde p^2\enspace ,\enspace a,b>0\enspace ,
\end{equation}
then the following inverse square root divergences appear on the I-V curves:

\begin{equation}
j_1,\,\mp j_4=\frac{\pm 1}{16\pi ^3}\tanh \left( \frac{|V|}{2T}\right)\!\!
\frac{\Theta \left( \pm \left( a-|V|\right) \right) }{\sqrt{\left| |V|
-a\right| }}\!\!\int\limits_{v_{x1}>0}\!\!\!dl\frac{v_{x1}}{v_{f1}}
B_{f,g1}(\hat {{\bf p}}_1)Re Q_{f,g2}(\hat {{\bf p}}_2)\frac D{\sqrt{b}}
\enspace \!,
\end{equation}

\begin{equation}
j_2,\,-j_3=\frac{1}{8\pi ^3}\tanh \left( \frac V{2T}\right) \frac{\Theta
\left( \pm \left( |V|-a\right) \right) }{\sqrt{\left| |V|-a\right| }}
\int\limits_{v_{x1}>0}\!\!\!\!dl\frac{v_{x1}}{v_{f1}}B_{f,g1}(\hat {{\bf p}
}_1)Re Q_{f,g2}(\hat {{\bf p}}_2)\frac D{\sqrt{b}}\enspace \!.
\end{equation}
In this section coefficients $B_{f2}$, $Q_{f2}^*$
originate from the corresponding expression for $f^+$ for second
superconductor, while $B_{f1}$, $Q_{f1}$ come from the expression for
$f$ for the first superconductor.

In the case of nonanalytical behavior of the function $h_2$ near some line
one has, for example,
\begin{equation}
\label{lbh}h_2=a+\left(b\Theta (\tilde p)+c\Theta (-\tilde p)\right)\tilde p
\enspace ,\enspace a>0 \enspace .
\end{equation}

Then the logarithmic divergences of $j_{1,4}$ occur at $|V|=a$:
\begin{equation}
j_1,\,j_4=\frac{\ln \left| |V|-a\right| }{8\pi ^4}\tanh \left( \frac{|V|}{2T
}\right) \!\int\limits_{v_{x1}>0}\!\!dl\frac{v_{x1}}{v_{f1}}DB_{f,g1}(\hat {
{\bf p}}_1)Re Q_{f,g2}(\hat {{\bf p}}_2)\left( \frac 1c-\frac 1b\right)\!
\enspace ,
\end{equation}

and the following jumps appear in $j_{2,3}$ at $|V|=a$:
$$
j_2,\,-j_3=\!\frac 1{8\pi^3 }\tanh \left( \frac V{2T}\right)
\int\limits_{v_{x1}>0}\!\!\!dl\frac{v_{x1}}{v_{f1}}DB_{f,g1}(\hat {{\bf p}
}_1)Re Q_{f,g2}(\hat {{\bf p}}_2)\left( \frac{\Theta \left( \left( |V|-a\right)
b\right) }{|b|}+
\right.
$$
\begin{equation}
\label{2}
\left.
\qquad\qquad\qquad\qquad\qquad\qquad
+\frac{\Theta \left( \left( a-|V|\right) c\right) }{|c|}
\right) \enspace .
\end{equation}

At last, let us consider the case of the line of extrema of the quantity $
\left| h_1- h_2\right|$:
\begin{equation}
\label{jh1}\left| h_1-h_2\right| =a\pm b\tilde p^2\enspace ,\enspace a,b>0
\enspace .
\end{equation}
Then one gets divergences on the I-V curves of the inverse square root
types:
\begin{equation}
j_{1},-j_{4}=\frac{\mp\Theta \left( \pm \left(
a-|V|\right) \right) }{16\pi ^3\,\sqrt{\left| |V|-a\right| }}\int
\limits_{v_{x1}>0}\frac{dl}{\sqrt{b}} N_{f,g} \enspace ,
\end{equation}

\begin{equation}
j_{2},-j_{3}=\frac{sgn(V)\Theta \left( \pm \left( |V|-a\right) \right)}{
8\pi^3\,\sqrt{\left| |V|-a\right| }}\int\limits_{v_{x1}>0}
\frac {dl}{\sqrt{b}}N_{f,g} \enspace ,
\end{equation}

where we introduce
\begin{equation}
N_{f,g}=\frac{v_{x1}}{v_{f1}} D Re\left(Q_{f,g1}Q_{f,g2}^*\right) \left| \tanh
\left( \frac{h_1}{2T}\right) -\tanh \left( \frac{h_2 }{2T}\right)
\right| \enspace .
\end{equation}

In the case when the quantity $\left| h_1-h_2\right| $ near some line
behaves like (\ref{lbh}), the logarithmic divergences of $j_{1,4}$ appear at
$|V|=a$:

\begin{equation}
j_{1},-j_{4}=\frac{\ln \left| |V|-a\right| }{8\pi ^4}\int
\limits_{v_{x1}>0}dl N_{f,g}
\left( \frac 1b-\frac 1c\right)\enspace ,
\end{equation}

and the following jumps take place in $j_{2,3}$ at $|V|=a$:
\begin{equation}
j_{2},-j_{3}=\frac{sgn(V)}{8\pi^3 } \int\limits_{v_{x1}>0}dl  N_{f,g}
 \left( \frac{\Theta \left( \left( |V|-a\right) b\right)
}{|b|}+\frac{\Theta \left( \left( a-|V|\right) c\right) }{|c|}\right)
\label{jh2}
\end{equation}
The results for the line of extrema of the quantity $h_1+h_2$ follow from
(\ref{jh1})-(\ref{jh2}) after the substitution $h_2\rightarrow -h_2$, $
j_1\rightarrow -j_1$, $Q_{f2}^*\rightarrow Q_{f2}$.
One should note, that in addition to the presented in this subsection
singular points there are the analogous ones which may be described by the
same equations (\ref{1})-(\ref{2}) after the interchange $1\leftrightarrow 2$.

The results of numerical calculations for $j_m(v)$ ($v=V/\Delta _0$) for the
case when there is surface pair breaking from both sides of the tunnel
barrier are shown in Figs. 5.1, 5.2. The tunnel junction between two
identical $d$-wave superconductors is considered for
the particular case of ''mirror'' junction, when the barrier is a
reflection-symmetry plane of the superconducting electrodes: $\Delta
_{1\infty }(\hat {\bf p}_1)=\Delta _{2\infty }(\check {\bf p}_1)=
\Delta _0\cos (2\phi-2\phi _o)$. As earlier we let (see Fig. 1)
$\phi _o=\pi /9$, $T=0.45T_c$, $\Delta _0/(2T)=2$, $D\propto \cos ^2(\phi )$.
There are the
inverse square root singular points on the curves $j_{1,2,3,4}$ at voltage $
V=2h_m=1.4\Delta _0$. For $V=2h_{ed}=1.26\Delta _0$ there are
logarithmic divergences of $j_1$, $j_4$ and jumps of $j_2$, $j_3$. At
voltage $V=0.635\Delta _0$ curves $j_1$, $j_2$ have kink-like behavior, which
may be associated with the interplay between the contributions of
mid-gap states from one
side of the junction and of
continuous quasiparticle spectrum from another side:
the function $\min \left( |\cos (2\phi -2\phi _o)|,|\cos (2\phi
+2\phi _o)|\right) $ has maximal value $0.635$ for the direction $\phi =\pm
0.79$ at which $B_f\neq 0$.

\section{Conclusions}

As we have shown above, there may be a large variety of different types of
nonanalytical points on the I-V curves for tunnel junctions in anisotropically
paired superconductors. The singular behavior differ essentially from that
which is characteristic for case of s-wave isotropic superconductors. The
quality of the barrier plane may have a great influence on the manifestations
of the singular points in question. For example, instead of the divergences
of the current, which are obtained for the idealized system, there are peaks,
whose practical finite magnitudes are sensitive to the values of
elastic and inelastic scattering processes, roughness of the barrier plane,
and even the finite value of the junction transparency. The inelastic
scattering processes are common for removing the singularities of the tunnel
current in s-wave isotropic and  in anisotropically paired superconductors.
The elastic scattering processes are pair breaking  just for the
anisotropically paired superconductors. Besides, these factors together with
the quality of the barrier plane and the finite value of transparency may
broaden the delta-peaks in quasiparticle density of states and,
subsequently, wash out the corresponding peaks in the tunnel current.
Nevertheless, the characteristic behavior of the I-V curves considered above
is observable under the certain realistic conditions and may be employed
as a sensitive test for identifying the anisotropic types of pairing in the
superconductors, in particular, different signs of the
order parameter on the Fermi surface.

\section*{Acknowledgements}
One of us (Yu.S.B.) is grateful to D. Rainer and J.A. Sauls for stimulating
discussions and for sending the preprints of their works before
publication. This work was supported by the grant No. 96-02-16249
of the Russian Foundation for Basic Research. A.A.S. acknowledges
Forschungszentrum J\"ulich for financial support.

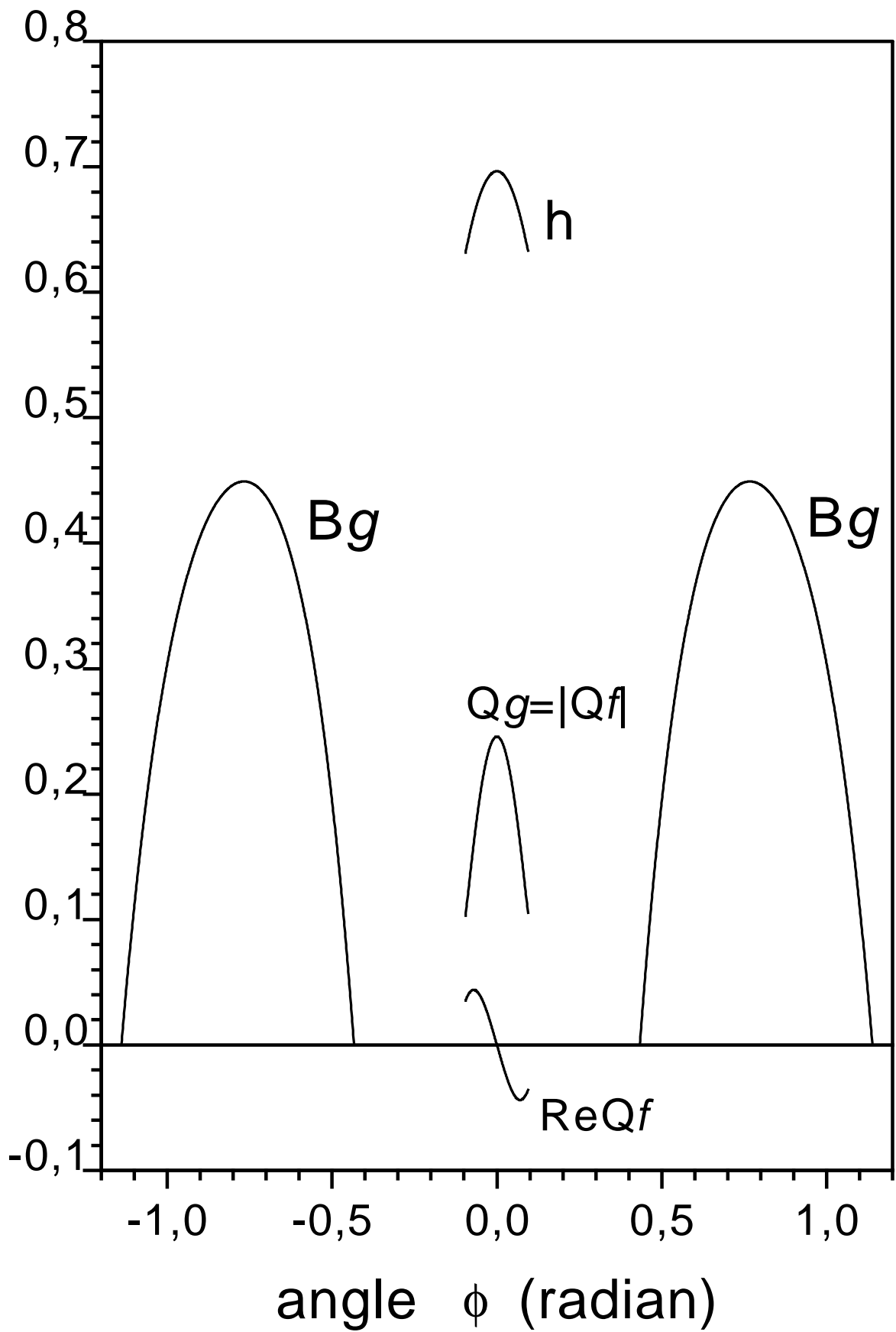
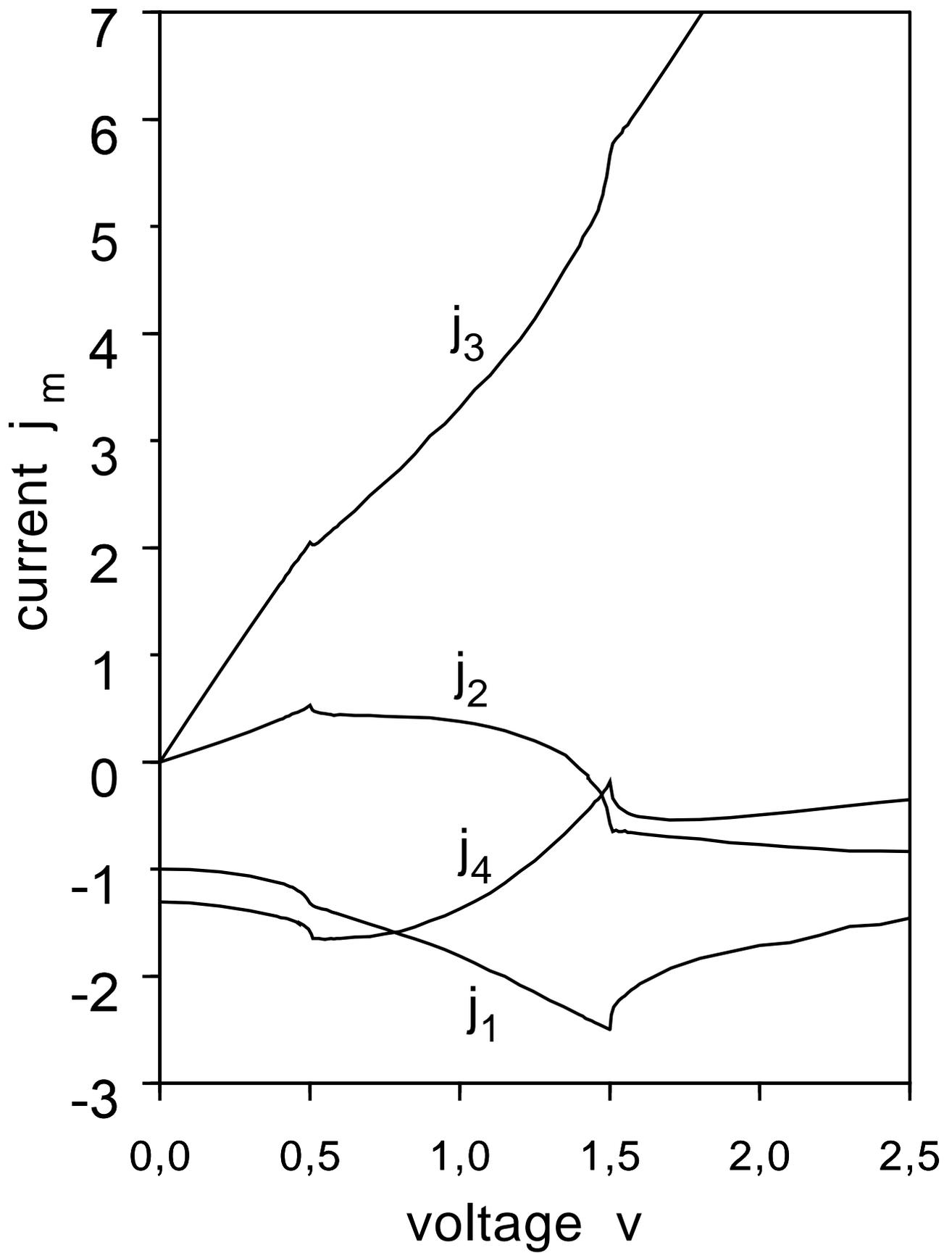
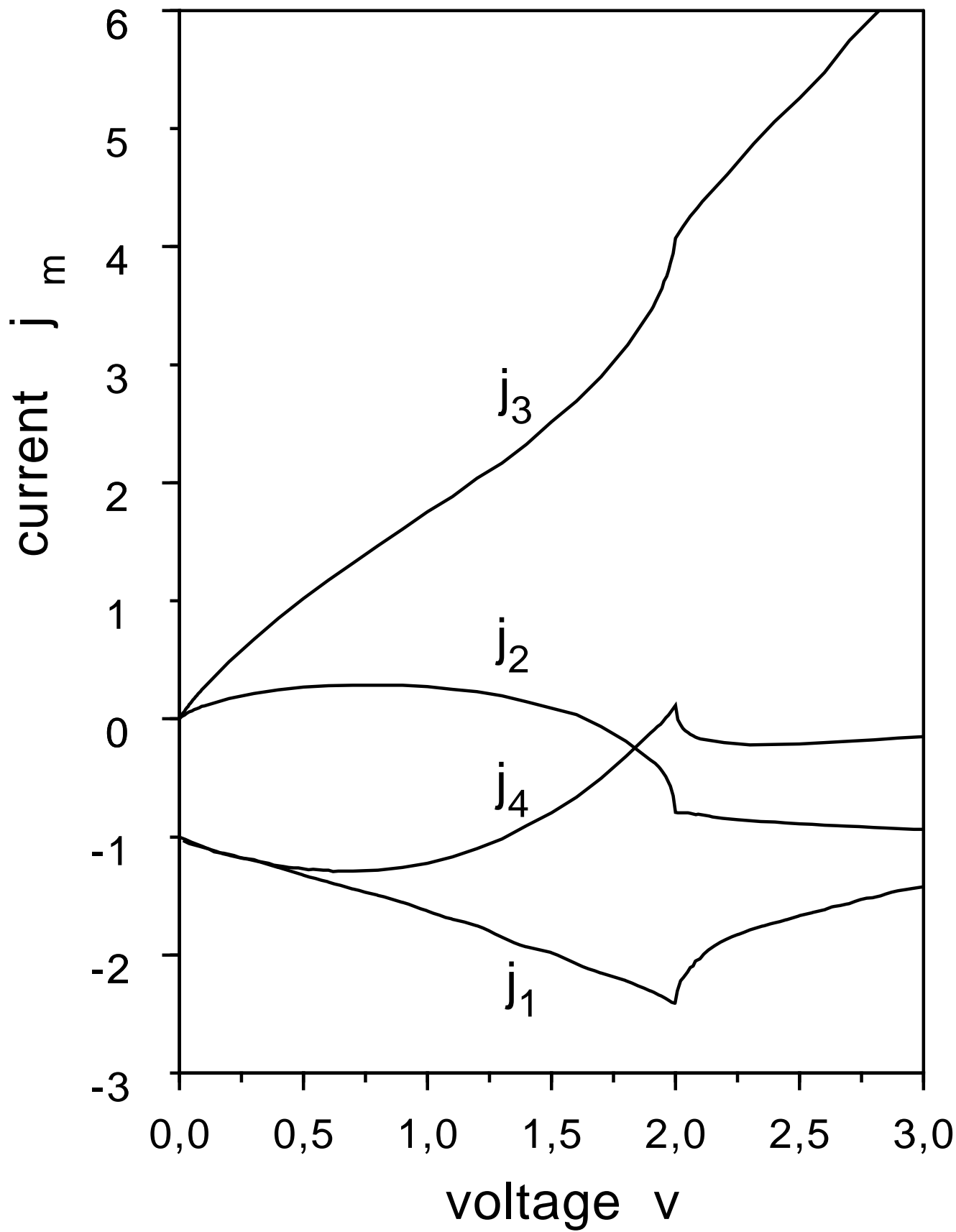
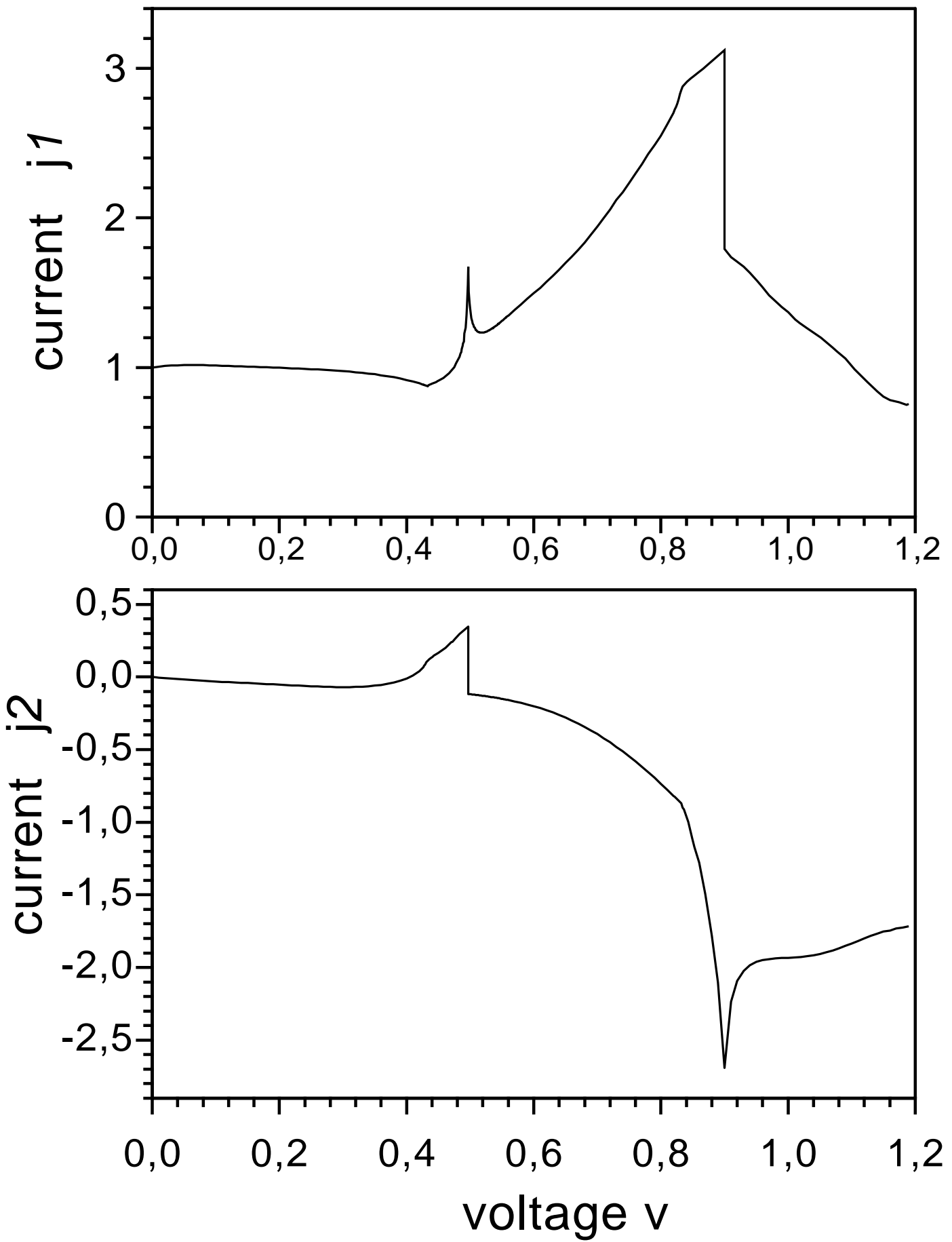
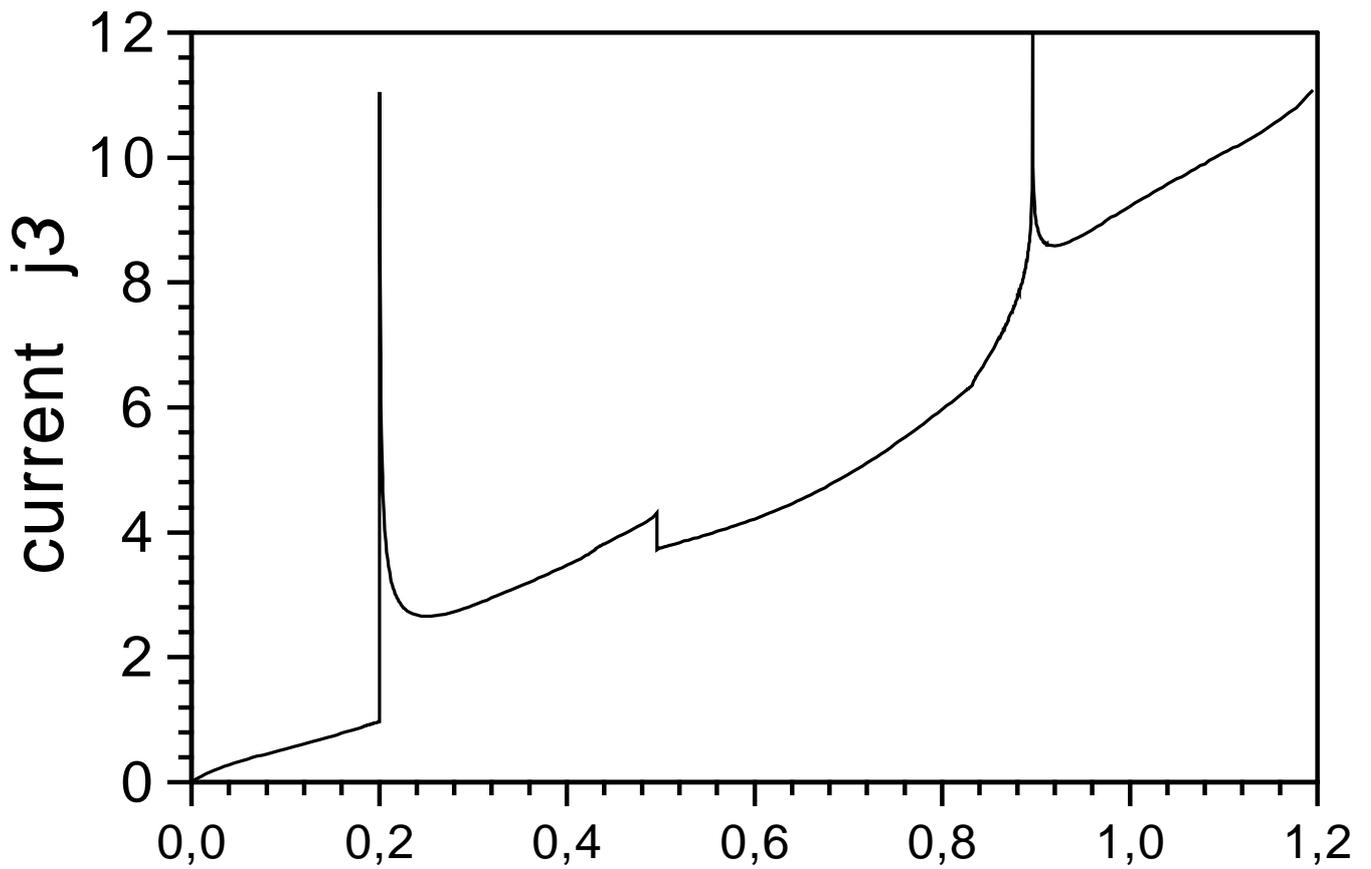
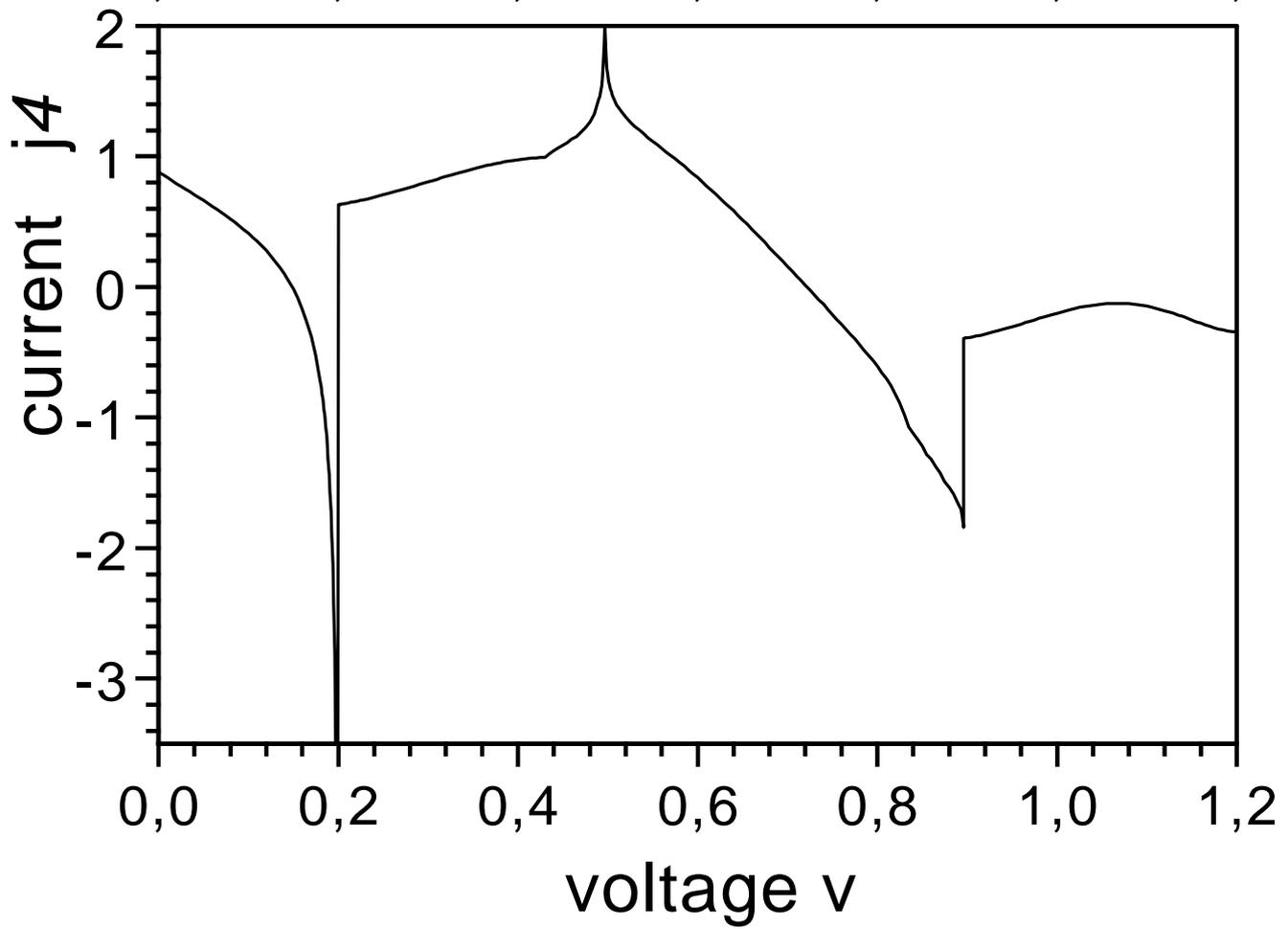
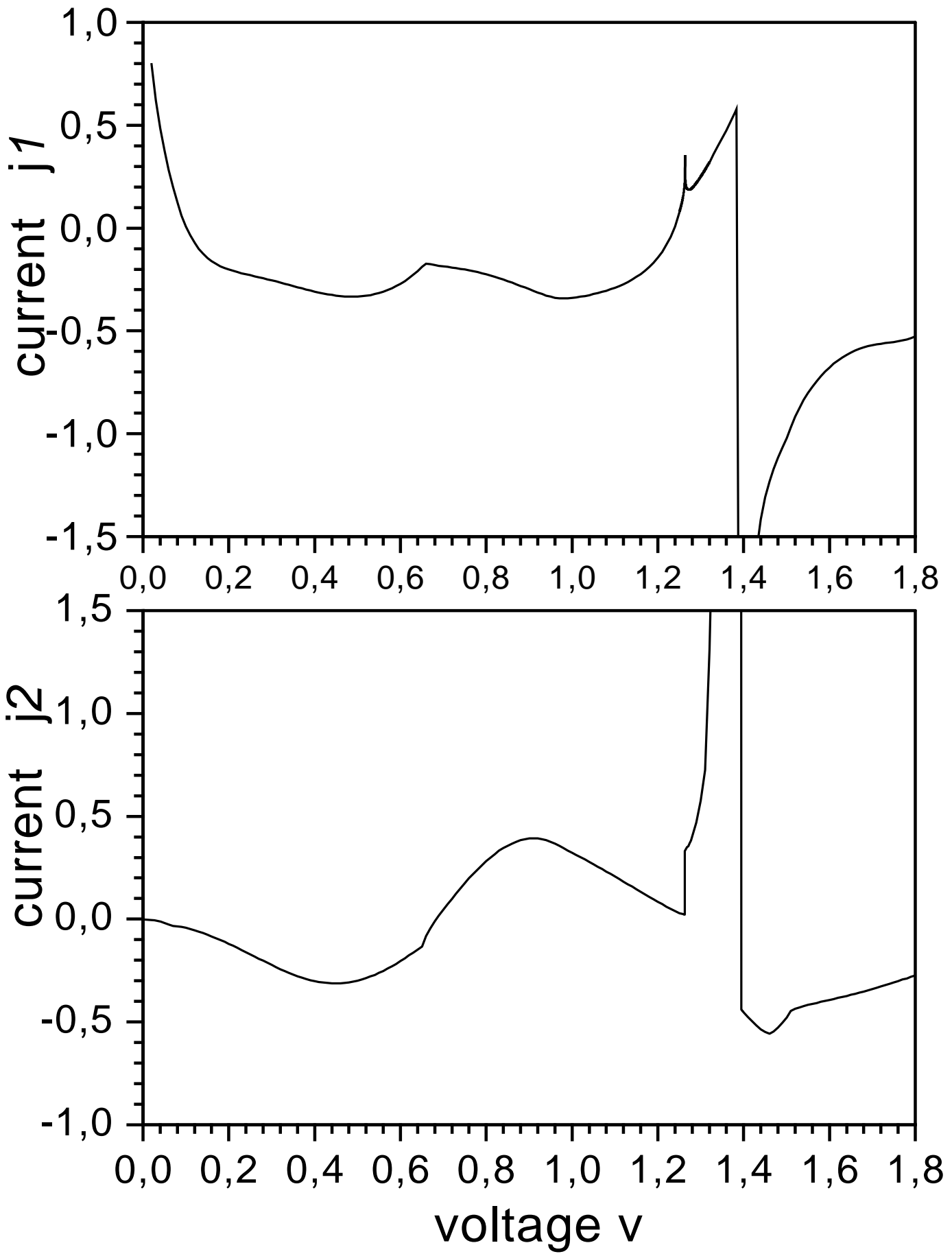
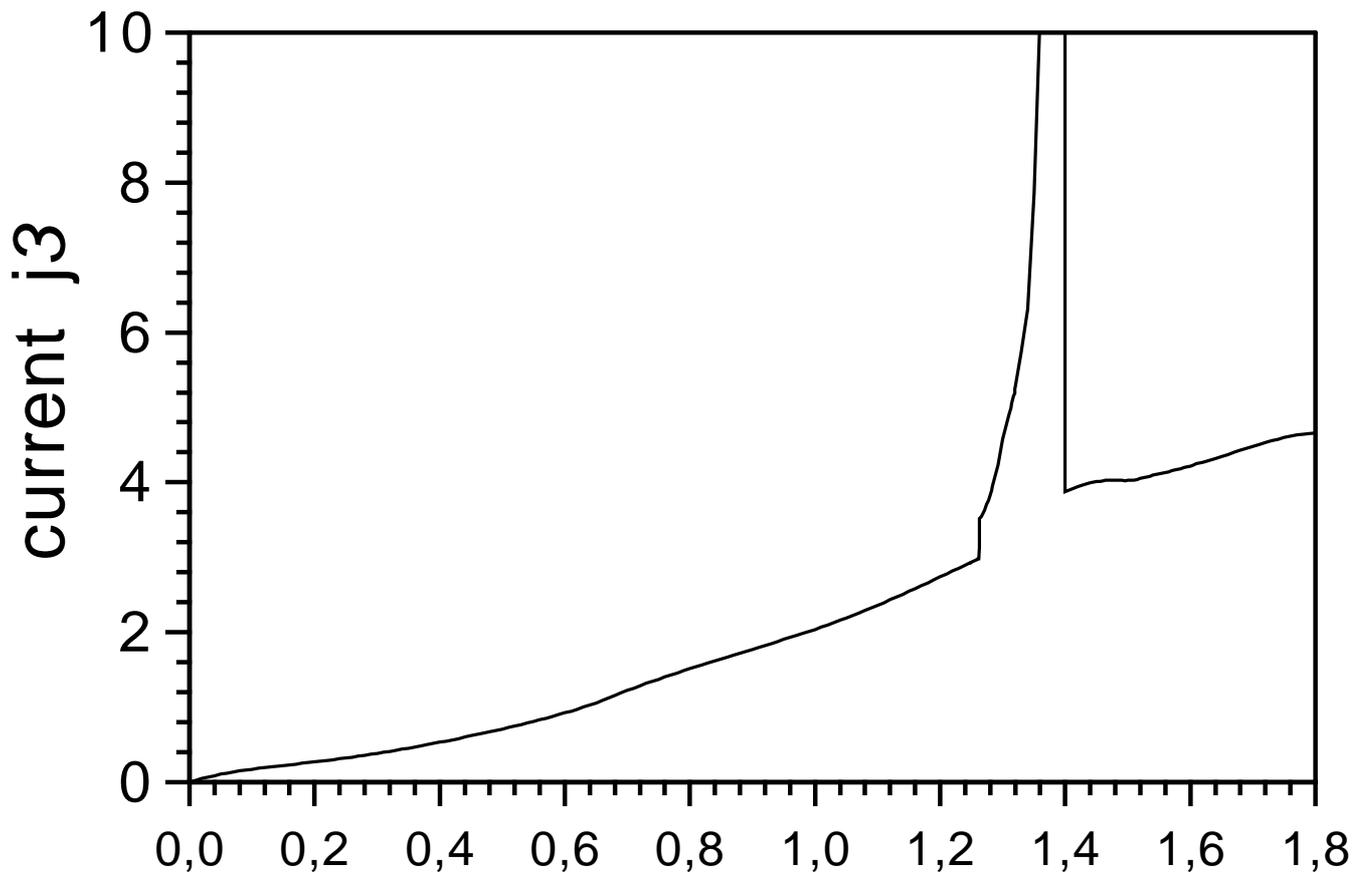
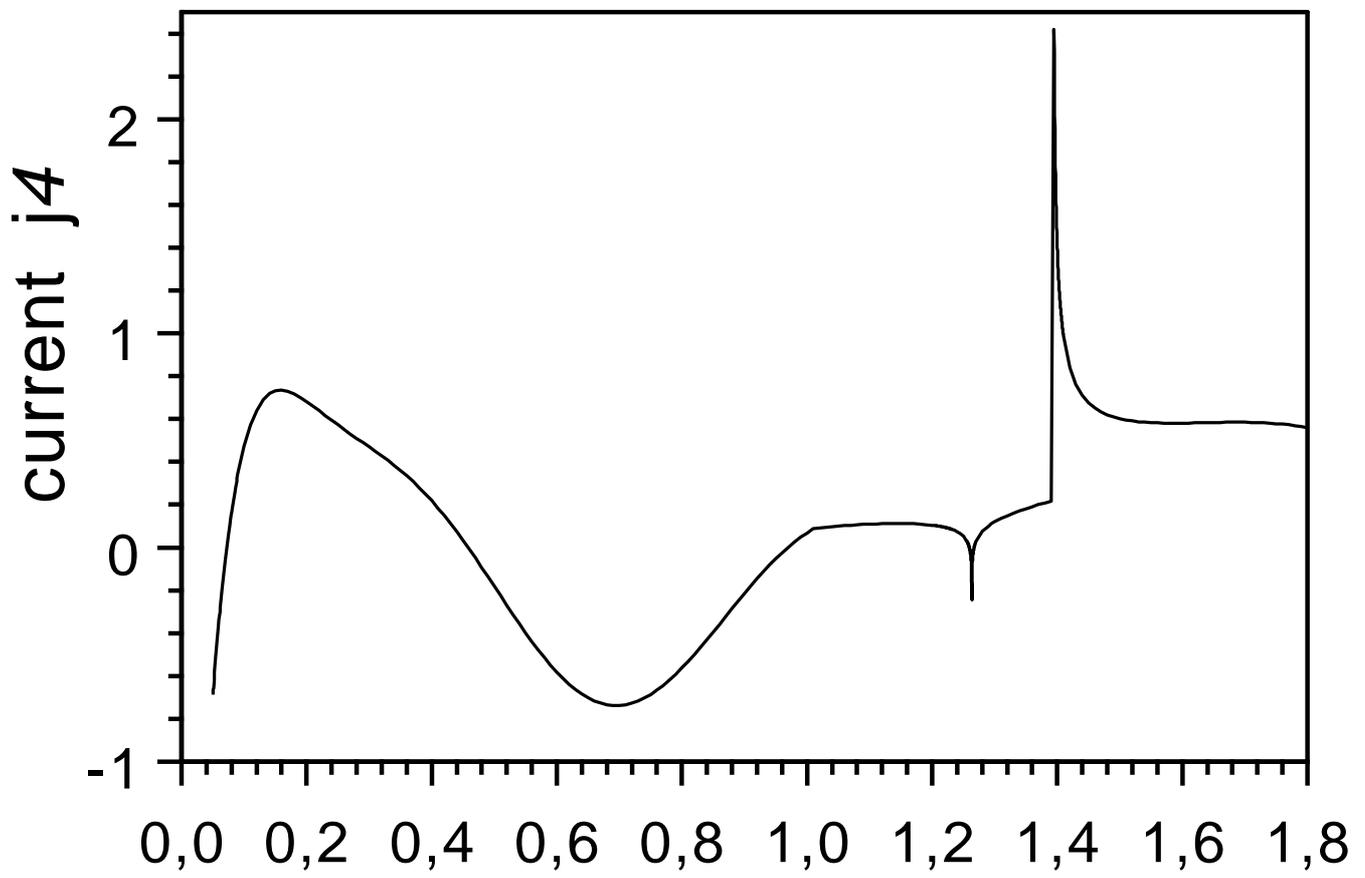
\end{document}